\documentclass[prd,preprint,showpacs,preprintnumbers,amsmath,amssymb,nofootinbib]{revtex4}

\usepackage{graphicx}
\usepackage{dcolumn}
\usepackage{bm}

\usepackage{t1enc}
\usepackage[english]{babel}   
\usepackage[latin1]{inputenc}

\newcommand{\pa}{\partial} 
\newcommand{\f}{\frac}

\newcommand{\no}{\nonumber\\}
\allowdisplaybreaks

\begin{document}

\title{Third post-Newtonian constrained canonical dynamics\\for binary point
masses in harmonic coordinates}

\author{Raoul-Martin Memmesheimer}
\email{rmm@chaos.gwdg.de}
\author{Gerhard Sch\"afer}
 \email{gos@tpi.uni-jena.de}
\affiliation{Theoretisch-Physikalisches Institut,
  Friedrich-Schiller-Universit\"at Jena\\
Max-Wien-Platz 1, 07743 Jena, Germany}

\date{\today}

\begin{abstract}
The conservative dynamics of two point masses given in harmonic coordinates up
to the third post-Newtonian (3pN) order is
treated within the framework of constrained canonical dynamics.
A representation of the 
approximate
Poincar\'e algebra
is constructed with the aid of Dirac brackets.
Uniqueness
of the generators of the Poincar\'e group
resp.\ the integrals of motion
is achieved by imposing 
their action on the point mass coordinates to be identical with that of the usual
 infinitesimal Poincar\'e transformations.
The second post-Coulombian approximation
to the dynamics of two point charges as predicted by
Feynman-Wheeler electrodynamics in Lorentz gauge is treated similarly.
\end{abstract}

                             
\maketitle                                         
\section{\label{sec:Intro}Introduction}

High post-Newtonian accurate description of general relativistic dynamics of
compact binaries in harmonic gauge has many applications in relativistic
astrophysics, notably in connection with binary pulsars and future
gravitational wave astronomy, see e.g.\ \cite{LBK04, Blanchet02, HR00}.
Inspiralling compact binaries are even the most promising sources to be
detected by ground-based interferometers such as LIGO, VIRGO and GEO600. The
corresponding higher order post-Coulombian approximation
offers a simpler analogue to
post-Newtonian dynamics
relevant also on its own \cite{Buonanno00, KS2001}.

The approximate analytical dynamics of compact binary systems in general
relativity is most often treated in harmonic coordinates (see \cite{Damour83}
and references therein). Quite recently the dynamics of binary point masses
has been completed to the third post-Newtonian (3pN) order \cite{BF00a, BF00b,
BF01a, BF01b, BDE04,IF03,Itoh04}. Hereby results derived by means of the canonical
formalism of Arnowitt, Deser and Misner have been confirmed
\cite{JS98,JS01e98,JS99,DJS00b,DJS01e00,DJS01a,DJS01b}. In approaches based on
the use of
harmonic coordinates, the dynamics was first obtained under the form of
ordinary second order 3pN equations of motion satisfied
by the particle trajectories.
The Lagrangian corresponding to the conservative part of the motion turns out to be of
higher order in the time derivatives of the point mass coordinates
\cite{ABF01}. This feature is shared by the Lagrangian of
Feynman-Wheeler
electrodynamics in Lorentz gauge derived by Kerner
\cite{FW49, Kerner62}. In both cases, Euler-Lagrange equations, of third order
or higher, admit a wide class of solutions including
physically irrelevant ones that do not reduce to the Newtonian solution
in the limit where the speed of light $c$ tends to infinity.
This can be seen directly from the number of independent initial
data including higher than first order time derivatives of the position variable. 
When demanding the Newtonian limit, it becomes possible to re-derive
the ordinary second order equations of motion by an iterative order reduction procedure.

The higher order property of these Lagrangians actually arises from the fact that
the dynamics in harmonic coordinates as well as in Lorentz gauge
for electromagnetism are approximately Poincar\'e invariant.
Indeed, the so-called no-interaction theorem by Martin and Sanz \cite{MS79} states that
Lagrangians of
point particles 
derived by means of a slow motion approximation from some classical field theory
must contain higher order derivatives from second order level in the
$1/c^2$ expansion, if
approximate manifest Poincar\'e invariance is maintained. For arbitrary approximately
Poincar\'e invariant point-particle dynamics,
higher order derivatives must be contained in the Lagrangian only from the third
order in powers of $1/c^2$ \cite{MS78}.
It can also be shown that, if exact Poincar\'e invariance of a system
with finitely many degrees of freedom is required, (i) Lagrangians including interactions 
must depend on time derivatives of infinite
order \cite{GKT87}, (ii) for point particle systems, the positions
may not be chosen as canonical
coordinates in Hamiltonian formalism (which reflects
the time non-locality due to retardation) \cite{CJS63,CJ64,Leutwyler64}.

For the approximate dynamics, appropriate contact transformations
lead to an ordinary Lagrangian but in a
non-harmonic grid \cite{DS91} resp.\ in a non-Lorentzian gauge. The transformed representation of
the dynamics can be described by means of an ordinary canonical formalism.
Another approach consists in constructing the canonical formalism
corresponding to the dynamics \emph{directly} in the original frame.

The original higher order Lagrangian is of singular type because of the
higher order derivatives occurring in ``small'' corrections i.e.\ in terms of higher order in
powers of $1/c^2$. The Hessian is thus multiplied with some positive power of
$1/c^2$. Therefore, it is non-invertible on
$\mathbb{R}[1/c^2]/(1/c^{2n+2})$, the ring of real polynomials in $1/c^2$
modulo $\mathcal{O}(1/c^{2n+2})$ we are working on at 
the $n$th order level of approximation.
Indeed, in expressing the highest order derivative as
a function of the others,
the Euler-Lagrange 
equations are multiplied with the inverse matrix of the Hessian, so that, in
particular,
the Newtonian part is multiplied with some power of $c^2$ and the Newtonian limit
does not exist anymore.
Independently from the $1/c^2$-power prefactor,
what we shall call the ``matrix part'' of the Hessian may be singular by itself,
leading to an additional singular
structure.

The canonical formalism for singular Lagrangians goes back to Dirac
\cite{Dirac50, Dirac58}, as well as Anderson and Bergmann \cite{AB1951}. The first
canonical treatment of dynamics derived from a slow motion approximation of
a classical relativistic theory described by a singular Lagrangian of higher
order (in the time derivatives) is due to Jaen, Llosa and Molina \cite{JLM86}.
Specializing on a class of approximate Lagrangians of a certain structure in
the $1/c^2$-power 
expansion  and demanding an invertible ``matrix part'' of the
Hessian,
they developed a method aimed at deriving an explicit expression for the
Hamiltonian as well as the Dirac brackets. They applied their formalism to the 2pC
dynamics in Lorentz gauge. However, the resulting Hamiltonian was not correct
because of computational errors. The first correct 2pC Hamiltonian is due to
Damour and Sch\"afer \cite{DS88}, their approach having been detailed in \cite{DS91}.
Later, Saito, Sugano, Ohta and Kimura proposed a method how to treat
general higher order
singular Lagrangians in canonical formalism.
They proved the
equivalence of
Lagrangian and Hamiltonian formulations
for singular Lagrangians
of higher order 
\cite{SSOK89, SSOK93}. A similar analysis has later been performed by Gr\`acia, Pons and
Rom\'an-Roy in a
geometrical framework \cite{GPR91}. The formalism given in \cite{SSOK89} was applied to a class
of 2pN Lagrangians to
which the post-Newtonian Lagrangians in harmonic coordinates do not belong.
It was also used by Ohta and Kimura for
investigating aspects of 2pC Feynman-Wheeler dynamics in Lorentz gauge \cite{OK93}.
Note that the approaches of articles \cite{SSOK89} and \cite{JLM86}
are crucially different. The singularity arising from the
 fact that we are working on the ring $\mathbb{R}[1/c^2]/(1/c^{2n+2})$
 is indeed not considered in Ref. \cite{SSOK89}.

 The aim of this paper is to formulate the canonical formalism for the
 conservative part of 3pN dynamics in harmonic coordinates as well as for
2pC  Feynman-Wheeler electrodynamics for two particles in
 Lorentz gauge and to analyze the
 dynamics in this framework.
For the formulation, we use a similar method
 as the one developed in Ref. \cite{JLM86} generalized to 3pN
 conservative binary dynamics.
For the first time we give the 3pN
 Hamiltonian in harmonic coordinates, the corresponding Dirac brackets, as
 well as a
 canonical representation of the Poincar\'e algebra of the 3pN and 2pC
 dynamics.
 
 The canonical description is always helpful for a better understanding of
 the dynamics. It is an extremely elegant tool to derive features such as
 symmetries and integrals of motion.
 The latter quantities, computed in
 harmonic coordinates at the 3pN order and specialized to the center of mass
 frame \cite{BI03}, are useful for the description of
 inspiralling compact binaries relevant as sources of gravitational waves.
 They allow the derivation of an analytic parametric ``generalized
 quasi-Keplerian'' solution to the 3pN accurate conservative equations of
 motion for compact binaries moving in eccentric orbits \cite{MGS04}. This is
 relevant in particular to construct post-Newtonian search templates for the detection of
 gravitational waves or to compare the numerical and post-Newtonian descriptions
 of such systems. Our integrals of motion
prove to be consistent with those computed by Andrade, Blanchet and Faye
\cite{ABF01} in Lagrangian formalism, providing a powerful cross
 check for the results.

Furthermore the inclusion of spin in post-Newtonian binary dynamics
using covariant spin supplementary condition also results in a dynamics that
is described by a higher order singular
Lagrangian or Hamiltonian when staying in harmonic coordinates
\cite{Damour82}.
Especially, the investigation of
its canonical description derived using the methods in the present
article will likely be useful for the
prediction of gravitational wave templates.

 This article continues work initiated by Stachel and Havas who derived
 in 1976 the Hamiltonians describing a class of dynamics including the
 1pN and 1pC ones and who computed the integrals of motion corresponding to the
 approximate Poincar\'e invariance \cite{SH76}. They announced a further
 article, where special interactions allowing the choice of the spatial
 coordinates as canonical coordinates were to be treated up to second order,
 but this article was never published. We do not follow the program they had
initially
 designed but rather concentrate on physically relevant
 interactions incompatible with the latter choice of canonical coordinates.

 The plan of the paper is as follows. In Sec.\  \ref{sec:2}, we
 outline the general constrained Lagrangian formalism for Lagrangians
 containing higher order time derivatives. We also show how to derive a full
 time-stable set of Lagrangian constraints not only for Lagrangians having a similar
 structure as in \cite{JLM86}, but also for cases where the ``matrix part'' of
 the Hessian is not invertible. In Sec.\  \ref{sec:3}, we outline the theory of
the corresponding Hamiltonian formalism. Sec.\  \ref{sec:4} is dedicated to a
 short description of the Poincar\'e algebra and its action on spatial
 coordinates. In Sec.\  \ref{sec:5} we apply the preceding results to the $3$pN dynamics of
 two point masses, showing the explicit Poincar\'e invariance and deriving the
 corresponding integrals of motion. The 2pC Lagrangian 
of Feynman-Wheeler electrodynamics for two charged point
 masses is treated similarly in Sec.\ \ref{sec:6}. Finally, in Sec.\ \ref{sec:7}, we summarize and
 discuss our results.

\section{\label{sec:2}Higher order singular Lagrangian point mass dynamics}
We start from the action integral of a higher order Lagrangian $L$ that does not depend
explicitly on time. It simply reads \cite{Borneas69, DS85}
\begin{equation}\label{e:2}
S \equiv \int_{t_0}^{t_1} dt L(q, q^{(1)},...,q^{(n)}),
\end{equation}
where $q$ is a short notation for the set of $f$ independent
variables $\{q_\mu\}$, $\mu=1,...,f$,
and where $q^{(i)}$ denotes the set
$\{q_\mu^{(i)}\}$ of their $i$th derivatives with
respect to time $t$. The highest order of time derivative
appearing in $L$
is denoted by $n$. From the action principle $\delta S = 0$,
we draw
the generalized Euler-Lagrange equations,
\begin{equation}\label{e:4}
\frac{\partial L}{\partial
q_\mu}
-\frac{d}{dt}\frac{\partial L}{\partial q_\mu^{(1)}}
+\frac{d^2}{dt^2}\frac{\partial L}{\partial q_\mu^{(2)}}
-\ldots + (-1)^n \frac{d^n}{dt^n}\frac{\partial L}{\partial q_\mu^{(n)}}=0,
\end{equation}
with $\mu=1,\ldots,f$, $d/dt$ denoting the total time derivative.
By 
collecting the terms of the $\mu$th equation that do not depend on $q^{(2n)}$
into a single function $f_\mu$ and isolating the highest order
derivatives $q^{(2n)}$, we may rewrite Eq.\ \eqref{e:4} as
\begin{equation}\label{e:5}
\sum_{\nu=1}^f q_{\nu}^{(2n)}\frac{\partial^2 L}{\partial q_{\mu}^{(n)} \partial
q_{\nu}^{(n)}}+f_{\mu}(q,...,q^{(2n-1)})=0.
\end{equation}
This relation shows
that the highest order 
time derivatives 
always occur multiplied with the Hessian matrix
\begin{equation}\label{e:6}
H_{\mu \nu}=\frac{\pa^2 L}{\pa q_{\mu}^{(n)} \pa
q_{\nu}^{(n)}},
\end{equation}                               
so that the Euler-Lagrange equations can be solved for
$q^{(2n)}$ as a function of the configuration space variables
$q,...,q^{(2n-1)}$ if and only if the Hessian is invertible. 

We now want to specialize to Lagrangians derived within slow-motion
approximation schemes of relativistic theories. These are for instance the
Lagrangian describing the conservative part of post-Newtonian dynamics that
has been determined up to third post-Newtonian order in Ref.\ \cite{ABF01}, or the
Lagrangian describing the $n$th post-Coulombian dynamics of two particles in
the Feynman-Wheeler theory of electromagnetism \cite{Kerner62}:
\begin{align}\label{e:7}
L(x, x^{(1)},...,x^{(n)})=&
-m_1 c^2 \bigg[1-\f{({x}_{1}^{(1)})^2}{c^2}\bigg]^{\f{1}{2}}
-m_2 c^2 \bigg[1-\f{({x}_{2}^{(1)})^2}{c^2}\bigg]^{\f{1}{2}}\no
&-e_1e_2 \sum_{k=0}^n \frac{1}{(2k)!}
\frac{(-D_1D_2)^k}{c^{2k}}\bigg[1-\f{({x}_{1}^{(1)} {x}_{2}^{(1)})}{c^2}\bigg]
 r^{2k-1}
+\mathcal{O}\left(\frac{1}{c^{2n+2}}\right),
\end{align}
where the letter $x$ is used to refer to configuration space variables
in order to emphasize the physical interpretation of $x_{ai}^{(j)}$ as
the $j$th derivative of the $i$th component ($i=1,2,3$) of the position vector of
the particle $a=1,2$; where $x_{a}^{(j)}$ stands for the $j$th time derivative
of the position vector of particle $a$, and $x^{(j)}=\{x_{a}^{(j)}\}$ 
for the set of all
configuration space variables that are time derivatives of $j$th order
($x\equiv x^{(0)}$).
Round brackets $(\,.\;\; .\,)$ indicate the scalar product, e.g.
$({x}_{1}^{(1)} {x}_{2}^{(1)})=\sum_{i=1}^3{x}_{1i}^{(1)} {x}_{2i}^{(1)}$;
if both vectors are identical we denote, e.g. $({x}_{1}^{(1)})^2$.  
We further introduced the operator $D_a$, which represents
the time derivative acting exclusively on the configuration space
variables belonging to the particle numbered $a$. $r$ 
is the absolute
value of the relative separation vector, and $e_a$ denotes the charge
of the particle $a$.
Note that, since the square
root may be expanded into a binomial series up to order
$1/c^{2n+2}$, 
it makes sense to
 consider $N$-particle
Lagrangians of the type \cite{JLM86}
\begin{equation}\label{e:8}
L=\frac{1}{2}\sum_{a=1}^N m_a (x_a^{(1)})^2+\sum_{s=0}^n \varepsilon^s
V_s(x,..., x^{(s)})+\mathcal{O}(\varepsilon^{n+1}),
\end{equation}                                     
with $\varepsilon\equiv\frac{1}{c^2}$.
Whereas for the 3pN dynamics there is actually no $x^{(3)}$ dependence, the
formalism can be adapted to this case (see Sec.\  \ref{sec:5}).
The Hessian of the above Lagrangian
\begin{equation}\label{e:9}
H_{ai\,bj}=\varepsilon^n\frac{\pa^2 V_n}{\pa x_{ai}^{(n)} \pa x_{bj}^{(n)}}
\end{equation}                      
is not invertible on the ring
$\mathbb{R}\left[\varepsilon\right]/\left(\varepsilon^{n+1}\right)$
on which the approximation scheme is defined. Thus, the Lagrangian is
singular and the system is subject to constraints which are now to be 
determined.

The primary Lagrangian constraints are given by all independent linear
combinations of the Euler-Lagrange equations that do not contain
$2n$th order time derivatives,
imposing thereby conditions
on the configuration space variables
(cf. e.g. \cite{Sundermeyer82, SM83}).
From Eq.\ \eqref{e:5} we see that they can be derived by
contracting the Euler-Lagrange equations with some null vectors of
the Hessian \eqref{e:9}, we shall refer to as $\lambda_r$. Let us first suppose that the ``matrix part''
$\frac{\pa^2 V_n}{\pa x_{ai}^{(n)} \pa x_{bj}^{(n)}}$ of $H_{ai\,bj}$ is
invertible \cite{JLM86}, a restriction we will skip later. Then, in our
approximation scheme, the null vectors of the Hessian are those that are
multiples of $\varepsilon$. The contraction of the Euler-Lagrange
equations with the canonical basis vectors of $\mathbb R^{3N}$ multiplied by $\varepsilon$ yields
all the primary constraints. In the notation of Eq.\
(\ref{e:5}) with the generalized coordinates being
the spatial coordinates $x$ and their time derivatives, these are given by
\begin{equation}\label{e:11}
\varepsilon f_{b j}(x,...,x^{(2n-1)})=\mathcal{O}(\varepsilon^{n+1}).
\end{equation}
Since $\frac{\pa^2 V_n}{\pa x_{a i}^{(n)}\pa x_{b j}^{(n)}}$ is invertible,
there are no more independent ones. Requiring the special form
\eqref{e:8} of the Lagrangian, the Euler-Lagrange equations read
\begin{gather}
  -m_\alpha x_{\alpha}^{(2)}+\sum_{s=0}^n \varepsilon^s A_{\alpha
    s}(x,..,x^{(2s)}) = \mathcal{O}(\varepsilon^{n+1}),\label{e:12}\\
A_{\alpha
    s}=\sum_{r=0}^s\left(-\frac{d}{dt}\right)^r\frac{\pa V_s}{\pa
    x_{\alpha}^{(r)}},\nonumber
\end{gather}
with $\alpha=ai$ and $m_a=m_\alpha$.
Now, by means of Eq.\ \eqref{e:12}, we rewrite the primary constraints \eqref{e:11} as
\begin{equation}\label{e:13}
\varepsilon m_\alpha x_{\alpha}^{(2)}=\varepsilon \sum_{s=0}^{n-1}\varepsilon^s
A_{\alpha s}(x,...,x^{(2s)})+\mathcal{O}(\varepsilon^{n+1}).
\end{equation}
We observe that they coincide up to the factor $\varepsilon$ with the
equations of motion of lower order in $\varepsilon$. Starting with the above equation,
we can derive a minimal stable set of constraints as explained in
appendix \ref{sec:A}, 
\begin{align}\label{e:21a}
x_{\alpha}^{(2+r)}&=\frac{1}{m_\alpha}\left[\sum_{s=0}^{n}\varepsilon^s
B_{\alpha,2+r,s}(x, x^{(1)})\right]
+\mathcal{O}(\varepsilon^{n+1}),
\end{align}
for $r=0,...,2n-3$. The precise definition of the functions $B_{\alpha,2+r,s}$
from the $A_{\alpha s}$'s is specified in the appendix.
It is worth noticing that the constraints corresponding to the case
$r=0$ agree with the equations of motion
after they have been 
iteratively reduced to order two
in the time derivatives
by removing
higher order time derivatives
with the help of the equations of motion of lower order in $\varepsilon$.
Similarly, the additional constraints agree with the appropriately reduced
time derivatives of the reduced equations of motion.
This justifies the preceding statement saying that the
constraints emerge by
requiring the Newtonian limit.

The ``matrix part'' $\frac{\pa^2 V_n}{\pa x_{ai}^{(n)} \pa
x_{bj}^{(n)}}$ of the Hessian of post-Newtonian Lagrangians linear in the
accelerations,
is not invertible. This can be cured
 by adding so-called double-zeros.
While this may change the rank of the
``matrix part'' of the Hessian, it does not influence the order-reduced
equations of motion \cite{BC80a,BC80b}. Because of the agreement between the
latter equations
and the constraints \eqref{e:21a}, established for an
invertible matrix $\frac{\pa^2 V_n}{\pa x_{ai}^{(n)} \pa x_{bj}^{(n)}}$, we
do not expect double-zeros to change the constraints either. A closer
investigation shows that this is indeed the case. We can state
even more generally that the expression for the constraints
and the construction of
the Hamiltonian remains unchanged
if the 
``matrix-part'' of the Hessian is non-invertible.\\

We thus suppose that $\frac{\pa^2 V_n}{\pa x_{ai}^{(n)} \pa x_{bj}^{(n)}}$ is
 not invertible and has a constant rank $3N-R$ all over the configuration
 space. Then, in the considered approximation scheme, all multiples of
 $\varepsilon$ are still null vectors of the Hessian, but there also $R$
 additional null vectors say $\lambda_{\rho\alpha}$, $\rho=1,...,R$, of order
 zero in $\varepsilon$. The primary constraints are obtained by contracting
 any of them with the equations of motion. We shall first consider the
 constraints emerging from the contraction of the Euler-Lagrange equations
 with the canonical basis vectors of $\mathbb R^{3N}$ times $\varepsilon$.
 Since the regularity of $\frac{\pa^2 V_n}{\pa x_{ai}^{(n)}\pa x_{bj}^{(n)}}$
 is actually not used in the derivation of the minimal time stable set
 they belong to, this set is again given by Eq.\
 \eqref{e:21a}. The primary constraints generated by the additional null
 vectors $\lambda_{\rho\alpha}$ read
\begin{align}\label{e:23}
\sum_\alpha \lambda_{\rho \alpha}(x,..., x^{(n)})\left[-m_\alpha
x_{\alpha}^{(2)}+\sum_{s=0}^{n}\varepsilon^s A_{\alpha
s}(x,..., x^{(2s)})\right]=&\mathcal{O}(\varepsilon^{n+1}),
\end{align}
where $\rho=1,..., R$ and where $\sum_\alpha$ denotes the sum over all pairs $\alpha=ai$
with $a=1,2$ and $i=1,2,3$;
$\lambda_{\rho \alpha}$ may depend on $x,...,x^{(n)}$ for the Hessian
itself possibly depends on these variables. We must now examine the additional
restrictions imposed by Eqs.\ \eqref{e:23} to the already derived constraint
surface given by the set of relations \eqref{e:21a}. On this surface, by
definition all higher order
derivatives $x^{(2)},..., x^{(2n-1)}$ entering Eqs.\ \eqref{e:23} may be
expressed by means of the coordinates
$x, x^{(1)}$ with the help of Eqs.\ \eqref{e:21a}. We find thus
\begin{align}
&\sum_\alpha \lambda_{\rho\alpha}(x,..., x^{(n)})\left[-m_\alpha
x_{\alpha}^{(2)}+\sum_{s=0}^{n} \varepsilon^s A_{\alpha
s}(x,...,x^{(2s)}) \right]+\mathcal{O}(\varepsilon^{n+1})\no
\approx &\sum_\alpha
\lambda_{\rho\alpha}(x, x^{(1)})\left[-m_\alpha
x_{\alpha}^{(2)}+\sum_{s=0}^{n}\varepsilon^s B_{\alpha, 2,
s}(x, x^{(1)})\right]+\mathcal{O}(\varepsilon^{n+1})
\label{e:24}
\underset{\eqref{e:21a}}
\approx \, \mathcal{O}(\varepsilon^{n+1}).
\end{align}
We emphasize the fact that these relations only hold on the constraint
 surface by using the weak equality symbol ``${\approx}$''. The system of equations
 \eqref{e:24} tells us that
 the constraints resulting from the additional null vectors of
 the Hessian are already fulfilled on the 
surface defined by the
 constraints \eqref{e:21a}. It is satisfied for all times due to 
 the time-stability 
property, so that the seemingly additional constraints
 are covered by the set \eqref{e:21a}. In short, the
 additional null vectors do not generate additional constraints. This fact
 enables us to perform the transition to Hamiltonian formalism regardless of
 the invertibility of the ``matrix part'' of the Hessian. Moreover, the Lagrangian
 constraints of the considered dynamics are still identical with the reduced
 equations of motion or their reduced derivatives. In particular, double zero
 terms, though they may change the rank of the Hessian, do not influence the formalism
 as long as the general structure \eqref{e:8} is maintained.
We observe that according to
above computation, unlike the usual theory, there are no arbitrary functions of time
emerging in the dynamics,
 even if the contraction of some of the additional null vectors with the
 Euler-Lagrange equations vanishes identically
 \cite{Sundermeyer82, SM83}. This is ultimately a consequence of
 the linear independence of the equations of motion at lowest order.

\section{\label{sec:3}Higher order singular canonical formalism}

A system described by a Lagrangian of higher order allows for a canonical
description with phase space variables $q^{(j)}$ and canonically conjugate momenta
$\Pi_j$ with $j=0,...,n-1$ \cite{Ostrogradski1850,Whittaker99}; 
$\Pi_j$ is the set of $j$th so-called
Ostrogradski momenta $\Pi_{j 1},...,\Pi_{j f}$
defined by
\begin{equation}\label{e:25}
\Pi_{j \mu}:=\sum^{n-j-1}_{k=0}\left(-\f{d}{dt}\right)^k\f{\pa L}{\pa
q_\mu^{(k+j+1)}},
\end{equation}
with $j=0,..., n-1$, $\mu=1,...,f$,
$n$ being the highest order derivative entering the Lagrangian
and $f$ the number of degrees of freedom.
For $j=0,..., n-2$ we may write alternatively
\begin{equation}\label{e:26}
\Pi_{j \mu}=\sum_{\nu=1}^f (-1)^{n-j-1} q_\nu^{(2n-j-1)}H_{\nu \mu}+K_{j \mu}(q,...,q^{(2n-j-2)})
\end{equation}
showing that the highest order time derivatives occur multiplied with
the Hessian. For $j=n-1$ we have
\begin{equation}\label{e:26a}
\Pi_{n-1\,\mu}=\frac{\pa L}{\pa q_\mu^{(n)}}.
\end{equation}
It is of the form \eqref{e:26} with $K_{j \mu}=0$ if $L$ is quadratic in
the $q^{(n)}_\mu$.
(This special case has not been accounted for in Ref. \cite{JLM86}.)
Let us first assume that the Hessian is regular. In this case,
the Ostrogradski transformation can be inverted by using an iterative
algorithm.
The implicit equation \eqref{e:26a} is locally solvable
for the $q^{(n)}$ and yields $q^{(n)}(q,...,q^{(n-1)},\Pi_{n-1})$.
Having computed the variables
$q^{(n+i)}$ 
with $i<j$,
we can invert the equation for the $n-j-1$th Ostrogradski momentum
for $q^{(n+j)}(q,...,q^{(n-1)},\Pi_{n-j-1},...,\Pi_{n-1})$.

The Hamiltonian of the system is
\begin{align}
H=&-L(q,...,q^{(n-1)},q^{(n)}(q,...,q^{(n-1)},\Pi_{n-1}))\\ \nonumber
&+\sum^{n-2}_{j=0}\sum_{\mu=1}^f \Pi_{j\mu}q^{(j+1)}_\mu
+\sum_{\mu=1}^f \Pi_{(n-1)\mu}q_\mu^{(n)}(q,...,q^{(n-1)},\Pi_{n-1}),
\end{align}
while the Hamiltonian equations of motion
take the familiar form
\begin{subequations}
\begin{eqnarray}
\label{e:27}\f{d}{dt} q_\mu^{(k)}&=&\f{\pa H}{\pa \Pi_{k\mu}},\\
\label{e:28}\f{d}{dt} \Pi_{k\mu} &=&-\f{\pa H}{\pa q_\mu^{(k)}},
\end{eqnarray}
\end{subequations}
with $k=0,...,n-1$.
Introducing the
(Ostrogradski-)Poisson bracket
\begin{equation}\label{e:28a}
\{F,G\} \equiv \sum_{j=0}^{n-1}\sum_{\mu=1}^f \f{\pa F}{\pa q_\mu^{(j)}}\f{\pa G}{\pa \Pi_{j
    \mu}} -\f{\pa G}{\pa q_\mu^{(j)}}\f{\pa F}{\pa \Pi_{j \mu}},
\end{equation}
the time evolution equations for a smooth function $F$ of the phase-space
variables and time $t$ takes the form
\begin{equation}\label{e:28b}
\frac{d}{dt}F=\{F,H\}+\f{\pa F}{\pa t}.
\end{equation}

Let us now turn to the case where the Hessian is singular. 
Then, the Ostrogradski transformation $(q,...,q^{(2n-1)})\rightarrow
(q,...,q^{(n-1)},\Pi_0,...,\Pi_{n-1})$ is not invertible anymore, or
equivalently, the phase space variables considered as functions of
$q,...,q^{(2n-1)}$ are not all independent. If the Hessian has rank
$f-r$ (for a maximum rank $f$), they are linked by $r$ independent relations. These are the
primary constraints
\begin{equation}\label{e:29}
\bar{\Psi}_a(q,...,q^{(n-1)},\Pi_0,...,\Pi_{n-1})=0
\end{equation}
following from the definition \eqref{e:25} of the momenta. To make sure that
the resulting constraint surface $\Gamma$
be a submanifold of phase space, we impose the ``regularity condition'',
demanding that zero be a regular value of $\bar{\Psi}$ regarded as a map
on phase space to $\mathbb{R}^{r}$.

The Hamiltonian $H$ as a function of the configuration space variables reads
\begin{align}\label{e:30}
H(q,...,q^{(2n-1)})=&-L(q,...,q^{(n)})+
\sum_{j=0}^{n-1}\sum_{\mu=0}^{f}\Pi_{j \mu}(q,...,q^{(2n-j-1)})q_\mu^{(j+1)}.
\end{align}
The remarkable fact is that the coordinates
$q^{(n)},...,q^{(2n-1)}$ appear only through the
combinations $\Pi_{j\mu}(q,...,q^{(2n-j-1)})$ due to the particular form
of the dependence of $L$ and $\Pi_{j \mu}$ on those coordinates.
Hence, $H$ actually depends only on
$q,...,q^{(n-1)}$ and $\Pi_0,...,\Pi_{n-1}$. This can be verified in a similar way
as in the absence of higher order derivatives \cite{HT92,Memmesheimer04}.
Therefore, we may view the Hamiltonian as a function of the phase space variables
$H = H(q,...,q^{(n-1)},\Pi_0,...,\Pi_{n-1})$ although it is not unique in the
case where the
Hessian is not invertible but defined modulo a linear combination of the
primary constraints $\sum_{a=1}^r c^a \bar{\Psi}_a$, with $c^a$ being functions
 of the phase-space coordinates \cite{HT92, Memmesheimer04}.
The time evolution of a smooth function $F$ of the phase space coordinates and
time is given by
\begin{equation}\label{e:32}
\frac{d}{dt}F=\{F,H\}+\sum_{a=1}^{r} u^a\{F,\bar{\Psi}_a\}+\f{\pa F}{\pa t},
\end{equation}
where $u^a$, $a=1,...,r$ are extra parameters,
and $\{\,.\,,.\,\}$
refers to the Poisson bracket \eqref{e:28a}. Additionally, at this level,
$\bar{\Psi}_a=0$ has to be imposed on the motion.
The relation $d\bar{\Psi}_a/dt=0$, $a=1,...,r$ may entail restrictions to the
$u^a$'s and/or lead to new, secondary, constraints. Using again the time
stability
property, we may
get further conditions, and so on. At the end we are left with a complete set of,
say, $K$, time stable constraints including the primary ones
\begin{equation}\label{e:34}
\Psi_k(q,...,q^{(n-1)},\Pi_0,...,\Pi_{n-1})=0,\, k=1,...,K,
\end{equation}
subject to the same regularity assumptions.

In constrained dynamics, there is an important distinction between two types of constraints:
(i) First class constraints are characterized by the property that their Poisson brackets
\eqref{e:28a} with all
the other constraints vanish on the constraint surface,
(ii) second-class constraints have at least one
non-vanishing Poisson bracket on $\Gamma$.
If the Poisson-bracket matrix $D$, defined in components as
$D_{kl}=\{\Psi_k,\Psi_l\}$ for $k,l=1,...,K$, has rank $A$ on $\Gamma$, then the
set of constraints \eqref{e:34} can always be linearly transformed into an
equivalent set
consisting of $K-A$ first-class constraints and
$A$ second-class constraints with 
invertible Poisson-bracket matrix on
$\Gamma$ \cite{HT92}. Conversely, if no first class constraints can be
obtained from Eqs.\ \eqref{e:34} by linear transformations,
the Poisson bracket-matrix $D$ is invertible on $\Gamma$.
In this case, the Dirac bracket of two phase space functions
may be defined by
\begin{equation}\label{e:34a}
\{F,G\}^*=\{F,G\}-\sum_{k,l=1}^{K}\{F,\Psi_k\}D^{-1}_{kl}\{\Psi_l,G\},
\end{equation}
where $D^{-1}$ is the inverse matrix of 
$D$, existing a least in a neighborhood of $\Gamma$.

The Dirac bracket keeps important features of 
the Poisson bracket; namely it is
bilinear, antisymmetric, acts as a derivation on each argument, and fulfills the
Jacobi identity. It actually is the
 restriction of the Poisson
bracket on the constraint surface \cite{MR99}. Finally, it has
by construction two additional important properties
\begin{align}
\label{e:34b}\{\Psi_k,F\}^*&=0,\\
\label{e:34c}\{G,F\}^*&\approx \{G,F\},
\end{align}
valid for arbitrary phase-space functions $F$ and for functions
$G$ of first class, i.e. functions whose Poisson bracket with any
constraint vanishes on $\Gamma$; 
``$\approx$'' represents the weak equality holding only on $\Gamma$.
We are thus allowed to simplify expressions
entering the Dirac bracket
by using the constraint equations, which amounts to setting the $\Psi_k$
to zero, before the final computation
of the bracket. With the help of such a tool we are in position to
reformulate the time
evolution equations equivalent to Eq.\ \eqref{e:32} on the constraint surface as
\begin{align}\label{e:34d}
\f{d}{dt}F=\{F,H_\text{r}\}^*+\f{\pa F}{\pa t},
\end{align}
where $H_\text{r}$ is the reduced Hamiltonian, derived from
$H=-L+\sum_{j=0}^{n-1}\sum_{\mu=0}^{f}\Pi_{j \mu}q_\mu^{(j+1)}$
by eliminating all the other coordinates in favor of the coordinates of the
constraint surface with the
help of the system's constraints \cite{SM83,JLM86,HT92,Memmesheimer04}.
The preceding canonical formalism has
been shown to be equivalent to the corresponding Lagrangian formalism in
Ref.\ \cite{SSOK89, GPR91}.

We now want to specialize on systems described by a
Lagrangian of the form \eqref{e:8}. From Eq.\
\eqref{e:25} we compute the Ostrogradski momenta,
\begin{gather}
\Pi_{j\alpha}=  m_\alpha
x_{\alpha}^{(1)}\delta_{j0}+\varepsilon^{j+1}
\Phi_{j\alpha}\left(x,...,x^{(2n-j-1)}\right) 
+O(\varepsilon^{ n+1}),
\no \label{e:35}
\Phi_{j\alpha}=\sum_{s=0}^{n-j-1}\varepsilon^s
\sum_{l=0}^s\left(-\frac{d}{dt}\right)^l\frac{\pa V_{s+j+1}}{\pa
  x_{\alpha}^{(l+j+1)}}.
\end{gather}
The Lagrangian constraints \eqref{e:21a} are already known.
Owing to the equivalence theorems proved in \cite{SSOK89, GPR91}, they can be
translated into constraints of the canonical formalism by keeping the first
$n-2$ relations and eliminating $x^{(2)},...,x^{(2n-1)}$ from the $n$
identities \eqref{e:35} by means of Eqs.\ \eqref{e:21a} (cf. Ref.\
\cite{JLM86}, in which
this method is applied whereas the equivalence had not yet been formally
stated). We find
\begin{subequations}
\begin{align}
 \label{e:37}
\omega_{r\alpha}&\equiv
x_\alpha^{(r)}-\frac{1}{m_\alpha}\sum_{s=0}^n
\varepsilon^s B_{\alpha,r,s}(x, x^{(1)})=O(\varepsilon^{n+1})\\
\label{e:36}\omega_{1\alpha}&\equiv x_\alpha^{(1)} -
\frac{1}{m_\alpha}\left[\Pi_{0\alpha}-\varepsilon 
\Phi_{0\alpha}(x,x^{(1)})\right]=O(\varepsilon^{n+1})\\
\label{e:38}
\chi_{j\alpha}&\equiv\Pi_{j
\alpha}-\varepsilon^{j+1}\Phi_{j\alpha}(x,x^{(1)})=O(\varepsilon^{n+1}),
\end{align}
\end{subequations}
with $r=2,...,n-1$ and $j=1,...,n-1$. The $\Phi_{j\alpha}(x, x^{(1)})$'s are
derived from the $\Phi_{j\alpha}\left(x,...,x^{(2n-j-1)}\right)$'s by
eliminating higher order derivatives with the help of Lagrangian constraints.
As the
constraints (\ref{e:37}-\ref{e:38}) obviously fulfill the regularity conditions, they
define a constraint surface $\Gamma$ that is a submanifold of phase space. As
coordinates, we may choose e.g.\
$x$ and $\Pi_{0}$, another possibility would be $x$ and $x^{(1)}$.

First class Hamiltonian constraints occur only if there are configuration
space variables that are arbitrary functions of time in Lagrangian formalism.
We have already seen that these are not present in our case so that there are
no first class constraints in the dynamics under consideration. This result
can also be proven by direct computation of the Poisson-bracket matrix (cf.
appendix \ref{sec:C} and \cite{JLM86}). It can be checked that
\begin{equation}\label{e:39}
D=\left(\begin{matrix}
        \{\chi_{k\alpha},\chi_{r\beta}\} & \{\chi_{k\alpha},\omega_{r\beta}\} \\
        \{\omega_{k\alpha},\chi_{r\beta}\} & \{\omega_{k\alpha},\omega_{r\beta}\}
        \end{matrix}\right)
\end{equation}
is indeed invertible, which implies
in particular that the number of constraints is even. Hence the constraint surface
$\Gamma$ has an even dimension, say $g$. Now, by an iterative procedure,
$D^{-1}$ may be expressed by means of submatrices of $D$ (cf. appendix \ref{sec:C}
and \cite{JLM86}). Adopting the splitting
\begin{equation}\label{e:40}
D^{-1}=\left(\begin{matrix}
        X & Y \\
        -Y^T & Z
        \end{matrix}\right),
\end{equation}
we may then write the Dirac bracket of two functions $f$ and $g$ on the phase
space as
\begin{align}
\{f, g\}^* =\{f,g\}
&-\sum_{k,l=1}^{n-1} \sum_{\alpha,\beta}
\{f,\chi_{k\alpha}\} X_{k\alpha\,,l\beta} \{\chi_{l\beta},g\}
-\sum_{k,l=1}^{n-1} \sum_{\alpha,\beta}
\{f,\chi_{k\alpha}\}Y_{k\alpha\,,l\beta} 
\{\omega_{l\beta},g\}
\no               
& 
+\sum_{k,l=1}^{n-1} \sum_{\alpha,\beta}
\{f,\omega_{k\alpha}\}Y_{\;l\beta\,,k\alpha}\{\chi_{l\beta},g\}
-\sum_{k,l=1}^{n-1} \sum_{\alpha,\beta}
\{f,\omega_{k\alpha}\}Z_{k\alpha\,,l\beta}\{\omega_{l\beta},g\}.
\label{e:41}
\end{align} 
This bracket defines a symplectic form on the constraint surface.
According to the theorem of Darboux, we can locally find canonically
conjugate coordinates 
$Q,P$ of $\Gamma$ 
such that it takes
the familiar shape 
\begin{equation}\label{e:42}                           
\{F, G\}^*=\sum_{\alpha=1}^{g/2} \frac{\pa F}{\pa Q_\alpha}\frac{\pa G}{\pa
P_\alpha}-\frac{\pa G}{\pa Q_\alpha}\frac{\pa F}{\pa P_\alpha}.
\end{equation}
We shall see in Sec.\ \ref{sec:5} and Sec.\ \ref{sec:6} that, for the
dynamics investigated in this article, it will even be possible to exhibit global
coordinates of this kind, which will greatly simplify computations involving Dirac brackets.
The global existence of canonically conjugate coordinates 
entails that the group of generalized canonical
transformations, i.e. the group of transformations that leave the Dirac
bracket invariant, is necessarily a subgroup of the
group of canonical transformations on the phase space \cite{SM83}.

\section{\label{sec:4} Poincar\'e algebra}

The symplectic structure on the phase space (resp. on the constraint surface) allows us
to endow
the vector space of scalar fields defined on $\Gamma$ with the structure of
 a Lie algebra. The Poisson-bracket (resp. Dirac-bracket) relations
satisfied by the generators of the infinitesimal (generalized) canonical
transformations corresponding to the action of a transformation group
are known to be identical to the Lie-bracket relations of
the generators of this group. In other words, there exists a
Lie algebra homomorphism between the Lie algebra of the transformation group
and that of the generators of the infinitesimal (generalized) canonical
transformations (see e.g. \cite{Marle98}).

Knowing the latter Lie algebra, we are enabled to reconstruct the
symmetry group locally. This can be done globally  only for simply connected
groups (unlike the Poincar\'e group). As we aim at determining the symmetry
group of some system in the phase space or on the constraint surface locally, 
it will be sufficient for us to consider the Lie algebra of the generators of infinitesimal
(generalized) canonical transformations (see e.g. \cite{SU00}).

By construction, the conservative pN dynamics in harmonic coordinates
as well as the pC dynamics of the Feynman-Wheeler theory in Lorentz gauge are approximately
Poincar\'e invariant.
Therefore, in a neighborhood of the
identity there exists an approximate
representation of the Poincar\'e group as a generalized canonical
transformation group on $\Gamma$. This means that on the constraint surface, there are generators
$H,P_i,J_i,G_i$ with $i=1,2,3$ of infinitesimal generalized
canonical transformations that approximately fulfill the Poincar\'e algebra with respect to
Dirac bracket: We have thus
\begin{subequations}
\begin{eqnarray}
\label{e:PA1} \{P_i,P_j\}^* & = & \mathcal{O}\left(\frac{1}{c^{2(n+1)}}\right),\\
\label{e:PA2} \{P_i,J_j\}^* &=& \sum_{k=1}^3\epsilon_{ijk} P_k+
\mathcal{O}\left(\frac{1}{c^{2(n+1)}}\right),\\
\label{e:PA3} \{J_i,J_j\}^* & = & \sum_{k=1}^3\epsilon_{ijk} J_k
+\mathcal{O}\left(\frac{1}{c^{2(n+1)}}\right),\\
\label{e:PA4} \{H, P_j\}^* & = & \mathcal{O}\left(\frac{1}{c^{2(n+1)}}\right),\\
\label{e:PA5} \{H, J_j\}^* & = & \mathcal{O}\left(\frac{1}{c^{2(n+1)}}\right),\\
\label{e:PA6} \{G_i, G_j\}^* & = & -\frac{1}{c^2}\sum_{k=1}^3\epsilon_{ijk}J_k
+\mathcal{O}\left(\frac{1}{c^{2(n+1)}}\right),\\ 
\label{e:PA7} \{G_i,H\}^*  & = &
P_i+\mathcal{O}\left(\frac{1}{c^{2(n+1)}}\right),\\
\label{e:PA8} \{J_i,G_j\}^*&=&\sum_{k=1}^3\epsilon_{ijk} G_k+
\mathcal{O}\left(\frac{1}{c^{2(n+1)}}\right),\\
\label{e:PA9} \{G_i,P_j\}^*&=&\frac{1}{c^2}\delta_{ij} H
+\mathcal{O}\left(\frac{1}{c^{2(n+1)}}\right),
\end{eqnarray}      
\end{subequations}                          
where $i,j=1,2,3$ are spatial indices and where $n$ is the order of approximation
($n=3$ for the 3pN, $n=2$ for the 2pC dynamics).
In the non-relativistic limit $c\rightarrow\infty$, the algebra
(\ref{e:PA1}-\ref{e:PA9})
reduces to the Galilean one.
In particular Eq.\ \eqref{e:PA9} becomes
\begin{equation} 
\label{e:GA9} \{G_i,P_j\}^*=\delta_{ij}M,
\end{equation}                          
$M$ being the total non-relativistic mass of the system.
Since all quantities are defined on the ring
$\mathbb{R}\left[1/c^2\right]/\left(1/c^{2(n+1)}\right)$,
strictly speaking,
the term $M c^2$ appearing in the generator
$H$ is not allowed. However, because it is a mere constant, it does not
change the dynamics  of the system. 
We may keep it for it suggests the physical interpretation of $H$ as
the total conserved energy. To harmonize the argument, we shall work with
a slightly modified form of Eq.\ \eqref{e:PA9}, namely
\begin{equation}\label{e:PA9a}\tag{\ref{e:PA9}'}
\{G_i,P_j\}^*=\delta_{ij}(M+\frac{1}{c^2}H)+\mathcal{O}\left(\frac{1}{c^{2(n+1)}}\right).
\end{equation}

The statement that a dynamics is
Poincar\'e invariant usually signifies
more than the existence of generators satisfying the Poincar\'e algebra. 
Namely, that the generators of the infinitesimal canonical invariance transformations
$H,P_i,J_i,G_i$ 
are the generators of the linear representation
of infinitesimal time translations, spatial
translations, spatial rotations and
Lorentz boosts acting on the particle components.
In short, the members of the Poincar\'e group
are to have the usual action on the physical system, when the phase space
coordinates $x(t)$ are interpreted as the positions of the
particles in Cartesian coordinates parameterized by the time $t$.
This is often emphasized by referring to this type of dynamics as
``manifestly'' Poincar\'e invariant.

In this perspective, we shall investigate one by one the actions of the
preceding generators on the phase space variables, acting on the coordinates
$x(t)$ regarded as the spatial coordinates of the system of $N$ point
particles at time $t$. The time evolution of $x^{(j)}(t), \Pi_{j}(t)$,
$j=1,...,n-1$ is given by the trajectories on the constraint surface; the
argument $t$ will be dropped below for the sake of simplicity. Throughout this
article we shall adopt the point of view of active transformations.

We start with the generator of infinitesimal time translations $H$.
By definition it generates a transformation
that can be interpreted, when regarded as active,
as a translation of the particles along their
trajectory. Now, the effect of an infinitesimal time translation about $\tau$ is
\begin{subequations}
\begin{eqnarray}
\label{e:43}\tau \f{d}{dt} x^{(k)}_{ai}=\delta_\tau x^{(k)}_{ai}&=&\tau\{
x^{(k)}_{ai},H\}^{*}+\mathcal{O}\left(\frac{1}{c^{2(n+1)}}\right),\\ 
\label{e:44}\tau \f{d}{dt}
\Pi_{k\,ai}=\delta_\tau \Pi_{k\,ai}&=&\tau\{\Pi_{k\,ai},H\}^{*}+\mathcal{O}\left(\frac{1}{c^{2(n+1)}}\right).
\end{eqnarray}
\end{subequations}
Note the use of Dirac brackets due to the fact that the trajectories are
restricted to $\Gamma$. From the equations of
motion \eqref{e:34d}, we see that $H$ may be chosen as the Hamiltonian or
the reduced Hamiltonian of the system as has been anticipated by the notation.
Since $H$ can be interpreted as the total energy of the system, the identity
$\{H,H\}^*=0$ traduces the conservation of energy in time.

The generators $P_i$ and $J_i$, interpreted as linear and angular momentum,
are the generators of infinitesimal spatial translations and
rotations. Even in absence of information on the dynamics,
we know their Poisson bracket with arbitrary scalar fields
defined on the phase space.
We know furthermore that generalized velocities are contravariant while
generalized momenta are covariant vectors.
Therefore, under an
infinitesimal translation about $\epsilon_i$ generated by $P_i$,
and an infinitesimal rotation about $\varphi_i$, generated by $J_i$, $i=1,2,3$,
the phase space coordinates representing the particle components transform as
\begin{align*} 
\epsilon_i&=\delta_{\epsilon}x_{ai}=
\sum_{j=1}^3
\epsilon_j\{x_{ai},P_j\},\\
 0&=\delta_{\epsilon}x_{ai}^{(l)}=
\sum_{j=1}^3
 \epsilon_j\{x^{(l)}_{ai},P_j\},\\
 0&=\delta_{\epsilon}\Pi_{m\,ai}=
 \sum_{j=1}^3\epsilon_j\{\Pi_{m\,ai},P_j\},\\
\sum_{j,k=1}^3
\epsilon_{ijk} \varphi_j x_{ak}^{(m)}&=\delta_{\varphi}x_{ai}^{(m)}=
\sum_{j=1}^3\varphi_j\{x_{ai}^{(m)},J_j\},\\
\sum_{j,k=1}^3
\epsilon_{ijk}\varphi_j  \Pi_{m\,ak}&=\delta_\varphi \Pi_{m\,ai}=
\sum_{j=1}^3 \varphi_j\{\Pi_{m\,ai},J_j\},
\end{align*}
 with $l=1,...,n-1$ and $m=0,...,n-1$. They lead to the differential equations
\begin{subequations}
 \begin{alignat}{2}
\label{e:48}\f{\pa P_i}{\pa \Pi_{m\,ak}}&=\delta_{ik}\delta_{0m}, \quad&
                     \f{\pa P_i}{\pa x_{ak}^{(m)}}&=0,\\
\label{e:49}\f{\pa J_i}{\pa \Pi_{m\,ak}}&=\sum_{j=1}^3\epsilon_{ijk}x^{(m)}_{aj}, \quad&
                     \f{\pa J_i}{\pa x_{aj}^{(m)}}&=\sum_{k=1}^3\epsilon_{ijk} \Pi_{m\,ak},
\end{alignat}
\end{subequations}
which fix the momenta $P_i$ and $J_i$ up to a constant. The Poisson bracket relations
of the these generators are well known; we have for instance
\begin{subequations}
\begin{eqnarray}
\label{e:51} \{P_i,J_j\} &=& \sum_{k=1}^3\epsilon_{ijk} P_k,\\
\label{e:52} \{J_i,J_j\} & = &\sum_{k=1}^3 \epsilon_{ijk} J_k.
\end{eqnarray} 
\end{subequations}                                                          
From Eqs.\ (\ref{e:48}-\ref{e:52}) we derive the unique result
\begin{eqnarray}
\label{e:53} P_i&=&\sum_{a=1}^{N}\Pi_{0\,ai},\\
\label{e:54} J_i&=&\sum_{m=0}^{n-1}\sum_{a=1}^{N}\sum_{j,k=1}^3 
\epsilon_{ijk}x^{(m)}_{aj}\Pi_{m\,ak}.
\end{eqnarray}
If a given dynamics is invariant under spatial translations and rotations, so
must be
the constraint equations and the constraint surface. In this case,
the generators of spatial translations
and rotations are thus first class functions. From Eq.\ \eqref{e:34c},
valid on the constraint surface, we see then that all Poisson-bracket relations
including $P_i$ and $J_i$ also hold on $\Gamma$ as Dirac-bracket relations.
This means that the momenta $P_i$ and $J_i$ displayed above are the generators
of infinitesimal spatial translations and rotations represented as
generalized canonical transformations on $\Gamma$.

 The physical interpretation of $G_i$ as generator of infinitesimal
Lorentz boosts allows us to determine its action on the particle
coordinates $x_{ai}$ \cite{SH76}. An infinitesimal boost
about $\upsilon_j$, $j=1,2,3$
acts on the space-time coordinates of a particle
following
\begin{subequations}
\begin{align}
\label{e:55}
\delta t_a&=-\frac{1}{c^2}\sum_{j=1}^3 \upsilon_j x_{aj},\\
\label{e:55a} \delta
x_{ai}&=-\upsilon_i t.
\end{align}
\end{subequations} 
A particle, located at $x_{ai}$ at time $t$, is located after the active
transformation at position
$x_{ai}'=x_{ai}+\delta x_{ai}$ at time $t_a'=t+\delta t_a$.
In the three-dimensional space, this results in
\begin{equation}\label{e:56}
x_{ai}'(t_a')=x_{ai}(t)+\delta x_{ai}=x_{ai}(t)-\upsilon_i t.
\end{equation}
Let us stress that, because we interpret the transformation as active, we keep
the same space-time coordinate system and simply boost the particles. We
choose the time coordinate $t$ to be eliminated in favor of $t_a'$ up to first
order in $\upsilon_i$. (However, since we are dealing with functional
identities valid for all the times, we could also proceed by substituting for
$t$.) Applying $t=t_a'+\frac{1}{c^2}\sum_{j=1}^3\upsilon_j x_{aj}(t)$ and
expanding Eq.\ \eqref{e:56} up to the linear order in $\upsilon_i$, we arrive
at the relation
\begin{eqnarray}
\label{iLT4} x_{ai}'(t_a')
&=& x_{ai}(t_a') +\frac{1}{c^2}\sum_{j=1}^3\upsilon_j
x_{aj}(t_a')\dot x_{ai}(t_a')
-\upsilon_i t_a' + \mathcal{O}(\upsilon^2).
\end{eqnarray}         
We have now expressed both sides of Eq.\ \eqref{e:56}
through the coordinate time $t_a'$. Since Eq.\ \eqref{iLT4}
is valid at any time, $t_a'$ is a mere ``dummy'' variable; it may be denoted
by $t$ again or even be dropped as a function argument.
This yields the following expression for the action of
 the generator $G_i$ on the spatial coordinates $x$ valid for any boost vector 
$\upsilon_i$
\begin{eqnarray}\label{e:61}
\sum_{j=1}^3 \upsilon_j
  \left(\frac{1}{c^2}x_{aj}\dot{x}_{ai}-\delta_{ij}t\right)=
\delta_\upsilon x_{ai}&=&\sum_{j=1}^3 \upsilon_j \{x_{ai},G_j\}^{*}
+\mathcal{O}\left(\frac{1}{c^{2(n+1)}}\right),
\end{eqnarray}
with $n$ being again the order of approximation.
Inserting the equations of motion \eqref{e:34d} into Eq.\ \eqref{e:61},
we deduce the
\emph{world line condition}
\begin{equation}\label{e:WB}
\{x_{ai},G_j\}^{*} = \f{1}{c^2} x_{aj}\{x_{ai},H\}^{*}-\delta_{ij}t
+\mathcal{O}\left(\frac{1}{c^{2(n+1)}}\right).
\end{equation}
This condition joined to the Poincar\'e algebra, to the expression of the
action of $P_i, J_i$ on the phase space variables, and to the physical
requirement that $H$ is the Hamiltonian of the system
will be sufficient to determine $G_i$ 
uniquely (up to arbitrary generalized canonical transformations).

In the following,
it will be useful to perform computations with the quantity $K_i=G_i+P_it$ instead of
$G_i$. Because of Eqs.\ (\ref{e:PA1}, \ref{e:PA2}, \ref{e:PA4}, \ref{e:PA9a}), $K_i$
fulfills the same Dirac-bracket relations as $G_i$, but is not explicitly time
dependent; it is not an integral of motion either.
Indeed,
we conclude from Eq.\ \eqref{e:PA7} combined with the conservation law
$\frac{d G_i}{dt}=0$ 
that  $\f{\pa G_i}{\pa t}=-P_i$.
Applying the definition of $K_i$ we find $\f{\pa K_i}{\pa t}=0$ and
$\{K_i,H\}^*   =  P_i$. The world-line condition for $K_j$ reads,
\begin{equation}\label{e:WBK}
\{x_{ai},K_j\}^{*} = \frac{1}{c^2}x_{aj}\{x_{ai},H\}^{*}
+\mathcal{O}\left(\frac{1}{c^{2(n+1)}}\right).
\end{equation}
It can be derived from Eq.\ \eqref{e:WB} and
from the first of the two equations \eqref{e:48}, which, 
due to the first class property of $P_i$ and relation \eqref{e:34c},
also holds as Dirac-bracket relation on the constraint surface
as mentioned above.

\section{\label{sec:5} Application to $3$pN dynamics}

The conservative part of $3$pN equations of motion for compact binaries
in harmonic coordinates allows for a Lagrangian of the form \cite{ABF01}
\begin{align}
L(x,x^{(1)},x^{(2)}) =& \frac{1}{2}\sum_{a=1}^2 m_a (x_a^{(1)})^2
+V_0(x) +
\frac{1}{c^2}V_1(x,x^{(1)}) \no
&+\frac{1}{c^4}V_2(x,x^{(1)},x^{(2)})
+\frac{1}{c^6}V_3(x,x^{(1)},x^{(2)})+
\mathcal{O}\left(\frac{1}{c^8}\right),
\end{align}
with $V_0(x)=G\frac{m_1 m_2}{r}$.
Beyond the fact it is restricted to a two-body system,
 it does not show any dependence on $x^{(3)}$
in the term $V_3$ by contrast to the general Lagrangian \eqref{e:8}.
The investigation of the dynamics can be performed in two different ways:
First, since the original formalism as displayed in sections
\ref{sec:2} and \ref{sec:3} does not require the invertibility 
of the Hessian, it may be applied on 
an artificially extended configuration space putting up however with 
lengthier computations. Second, we may
 consider
$V_3(x,x^{(1)},x^{(2)})$
as a higher order correction to $V_2(x, x^{(1)}, x^{(2)})$,
gathering both functions into a new quantity,
\begin{equation}
\bar V_2(x, x^{(1)}, x^{(2)})=V_2(x, x^{(1)}, x^{(2)})
+\f{1}{c^2}V_3(x, x^{(1)}, x^{(2)}).
\end{equation}
In this approach we write the 3pN Lagrangian (substituting $\varepsilon\equiv\frac{1}{c^2}$)  as
\begin{align}
L_{\text{3pN}}=&\frac{1}{2}\sum_{a=1}^2 m_a (x_a^{(1)})^2+V_0(x)+\varepsilon V_1(x,x^{(1)})
+\varepsilon^2 \bar V_2(x, x^{(1)}, x^{(2)})+\mathcal{O}\left(\varepsilon^4 \right),\label{e:L3pN}
\end{align}
without affecting the original accuracy. The formalism described
in sections \ref{sec:2} and \ref{sec:3} has to be appropriately modified.

It has been checked explicitly that both methods
lead to equivalent constraints,
to identical expressions for the elementary Dirac-brackets and to identical
conserved quantities on the constraint surface. 
However, since the second method is more adapted to the problem,
it demands less computational effort. Furthermore it
can be employed
for the computations of 2pC-dynamics when the terms of third order
are neglected.
We shall therefore use it for the subsequent calculations.

The equations of motion for the Lagrangian \eqref{e:L3pN} read
\begin{gather}
-m_\alpha x_{\alpha}^{(2)}+\sum_{s=0}^2 \varepsilon^s
A_{\alpha s}(x,...,x^{(2s)})=\mathcal{O}(\varepsilon^{4}),\label{e:3pNeqm}\\
A_{\alpha 0}=\frac{\pa V_0}{\pa x_{\alpha}},\,
A_{\alpha 1}=\frac{\pa V_1}{\pa x_{\alpha}}
-\frac{d}{dt}\frac{\pa V_1}{\pa x_{\alpha}^{(1)}},\,
A_{\alpha 2}=\frac{\pa \bar V_2}{\pa x_{\alpha}}-\frac{d}{dt}\frac{\pa
\bar V_2}{\pa x_{\alpha}^{(1)}}
+\frac{d^2}{dt^2}\frac{\pa \bar V_2}{\pa x_{\alpha}^{(2)}}\nonumber.
\end{gather}                                                               
The computation of the
constraints for a system described by the above Lagrangian is performed in
appendix \ref{sec:B}. The main difference
to the formalism exposed in Sec.\ \ref{sec:2}
is that the equations of motion have to be employed twice at the end
to guarantee time stability.
Finally we obtain a minimal set of time stable Lagrangian
constraints \eqref{e:pZB10}
\begin{equation} \label{e:pZB}
m_\alpha  x_\alpha^{(r)} - \sum_{s=0}^2\varepsilon^s B_{\alpha,r,s}(x,
x^{(1)})=\mathcal{O}(\varepsilon^4),
\end{equation}
for $\alpha=a\,i$
and $r=2,3$.
The Ostrogradski momenta derived from \eqref{e:35} are given by
\begin{subequations}
\begin{eqnarray}
\Pi_{0\alpha}&=&m_\alpha x_\alpha^{(1)}+\varepsilon
\Phi_{0\alpha}(x,...,x^{(3)})+\mathcal{O}(\varepsilon^4), \\ 
\Pi_{1\alpha}&=&\qquad \quad
\;\;\; \varepsilon^2\Phi_{1\alpha}(x,x^{(1)},x^{(2)})+\mathcal{O}(\varepsilon^4),
\end{eqnarray}
\end{subequations}
where the $\Phi_{0\alpha}$, $\Phi_{1\alpha}$ are now constructed with $V_1$
and $\bar V_2$:
\begin{subequations}
\begin{eqnarray}
\Phi_{0\alpha}&=&\frac{\pa V_1}{\pa x_\alpha^{(1)}}+\varepsilon  \frac{\pa
  \bar V_2}{\pa x_\alpha^{(1)}}
-\varepsilon \frac{d}{dt} \frac{\pa \bar V_2}{\pa x_\alpha^{(2)}}\\
\Phi_{1\alpha}&=& \frac{\pa \bar V_2}{\pa x_\alpha^{(2)}}.
\end{eqnarray}
\end{subequations}
The transformation of the Lagrangian constraints
into constraints on the phase space coordinates yields the Hamiltonian constraints
\begin{subequations}
\begin{align}\label{e:chi}
\chi_{1\alpha}& \equiv \Pi_{1\alpha}-\varepsilon^{2}\Phi_{1\alpha}(x,x^{(1)})=
\mathcal{O}(\varepsilon^4),\\
\label{e:omega} 
\omega_{1\alpha}&\equiv x_\alpha^{(1)}-\frac{1}{m_\alpha}[\Pi_{0\alpha}-\varepsilon
\Phi_{0\alpha} (x, x^{(1)})]=\mathcal{O}(\varepsilon^4).
\end{align}
\end{subequations}
The $\Phi_{j\alpha}(x, x^{(1)})$'s are derived from the
$\Phi_{j\alpha}\left(x,..., x^{(3-j)}\right)$'s by eliminating higher order
derivatives by means of Eqs.\ \eqref{e:pZB}. The Poisson-bracket matrix $D$
of the constraints (\ref{e:chi}, \ref{e:omega}) is regular, as shown
in appendix \ref{sec:C} by an explicit computation of $D^{-1}$. It is thus
possible to endow phase space with the Dirac brackets \eqref{e:34a}. Their
explicit expression can be found at the end of appendix \ref{sec:C}.

We are now in position to give the representation of the generators
of the Poincar\'e group as generators of infinitesimal generalized canonical transformations
with respect to the Dirac bracket for the conservative
3pN binary dynamics in harmonic coordinates. The phase space variables ${x}_1$ and ${x}_2$ are
the positions of the two point masses. The
generators act as usual on them and on their conjugate
momenta,
in the way specified in Sec.\ \ref{sec:4}.

 The reduced Hamiltonian, generator of infinitesimal time translation, can be computed from
\begin{equation}\label{e:30a} 
H_{3\text{pN}}=-L_{3\text{pN}}+\sum_{a=1}^2\sum_{i=1}^3\Pi_{0\,ai}x_{ai}^{(1)}+
\sum_{a=1}^2\sum_{i=1}^3\Pi_{1\,ai}x_{\,ai}^{(2)},
\end{equation}                                              
by using
the constraints
to eliminate $x^{(1)}$, $x^{(2)}$ and $\Pi_{1}$
in favor of the coordinates of the constraint surface. The latter are chosen to be
$x$ and $\Pi_0$
because of their ``Hamiltonian'' character
(a possible alternative is $x$ and $x^{(1)}$). In this grid,
the reduced 3pN Hamiltonian explicitly reads
\begin{align}
H_{\text{r},3\text{pN}}=&\,^0\!H_\text{N}+\frac{1}{c^2}\,^1\!H_\text{N}+
\frac{1}{c^4}\,^2\!H_\text{N}+\frac{1}{c^6}\,^3\!H_\text{N},\label{H3pN}\\[7pt]
\,^0\!H_\text{N}=&\f{{{ \Pi_{01}}}^2}{2m_1}+\f{{{ \Pi_{02}}}^2}{2m_2}-G\f{m_1
m_2}{r},
\nonumber\\[7pt]
\,^1\!H_\text{N}=&
-{\frac {{{\Pi_{01}}}^{4}}{8{{ m_1}}^{3}}}
-{\frac {{{\Pi_{02}}}^{4}}{8{{ m_2}}^{3}}}
+
{\frac {G}{2r}}
\left[7\left({ \Pi_{01} }{ \Pi_{02}} \right)
-{\frac{3 m_1}{{ m_2}}}{{\Pi_{02}}}^{2}
-{\frac{3 m_2}{{ m_1}}}{{\Pi_{01}}}^{2}
  +
{ } \left( { \Pi_{01}}n_{12} \right) { } \left( { \Pi_{02}}n_{12}
 \right)
 \right. \biggl.\biggr]
 \quad \nonumber\\
&
 +{\frac {{G}^{2}{ m_1}{ m_2} \left( { m_1}+{ m_2} \right)}{2{r}^{2}}},
\nonumber\\[7pt]
\,^2\!H_\text{N}=&
{\frac {{{ \Pi_{01}}}^{6}}{16 {{ m_1}}^{5}}}
+
{\frac {G}{16 {{ m_1}}^{2}{{ m_2}}^{2}\,r}}
\left[
{\frac {10{{ m_2}}^{3}}{{ m_1}}}{{ \Pi_{01}}}^{4}-15{ m_1}{
 m_2}{{ \Pi_{02}}}^{2}{{ \Pi_{01}}}^{2}
 +14{ m_1}{ m_2}
 \left({ \Pi_{02}}n_{12}\right)^{2}{{ \Pi_{01}}}^{2}
  \right. \nonumber\\ &\left.       \quad
-4{{ m_2}}^{2} \left( { \Pi_{02}}n_{12} \right)
 \left( { \Pi_{01}}n_{12} \right) {{ \Pi_{01}}}^{2}
  +4{{ m_2}}^{2}
\left( { \Pi_{01}}{ \Pi_{02}} \right) {{ \Pi_{01}}}^{2}
-2{ m_1}{m_2} \left( { \Pi_{01}}{ \Pi_{02}}
 \right) ^{2}
  \right. \nonumber\\ &\left.        \quad
 -12{ m_1}{ m_2} \left( { \Pi_{02}}n_{12}
 \right)  \left( { \Pi_{01}}n_{12} \right)  \left( {
\Pi_{01}}{ \Pi_{02}} \right)
 -3{ m_1}{ m_2}
 \left( { \Pi_{01}}n_{12}  \right) ^{2} \left( {
 \Pi_{02}}n_{12}\right) ^{2}
\right. \biggl.\biggr]\nonumber\\
&+
{\frac {{G}^{2}}{8{ m_1}{ m_2}\,{r}^{2}}}
\bigg[
22{{ m_2}}^{3}{{ \Pi_{01}}}^{2}+47{ m_1}{{ m_2}}^{2}{{
\Pi_{01}}}^{2}-4{{ m_2}}^{3}\left( { \Pi_{01}}n_{12}
 \right) ^{2}
  -70{{ m_1}}^{2}{ m_2} \left( {
 \Pi_{01}}{ \Pi_{02}} \right) 
  \nonumber\\ & \quad
 +16{{ m_1}}^{2}{ m_2}
 \left( { \Pi_{01}}n_{12} \right)  \left( { \Pi_{02}}n_{12} \right)
   -13
{ m_1}{{ m_2}}^{2} \left( { \Pi_{01}}n_{12}
\right) ^{2}
\bigg]
-
{\frac {{G}^{3}{ m_1}{ m_2}}{8{r}^{3}}}
\bigg[
{19}{ m_1}{ m_2}
\no&\quad
+4{{ m_1}}^{2}
\bigg]
+1\longleftrightarrow 2, \nonumber\\[7pt]
\,^3\!H_\text{N}=&-{\frac {5 {{ \Pi_{01}}}^{8}}{128 {{ m_1}}^{7}}}+
{\frac {G}{32 {{ m_1}}^{2}{{ m_2}}^{2}\,r}}
\bigg[
-{\frac {14 {{ m_2}}^{3}{{ \Pi_{01}}}^{6}}{{{ m_1}}^{3}}}
+{\frac{58{ m_2}}{{ m_1}}}{{ \Pi_{02}}}^{2}{{ \Pi_{01}}}^{4}
\no&\quad
+{
\frac {28{{ m_2}}^{2}}{{{ m_1}}^{2}}} \left( { \Pi_{01}}n_{12} \right) 
 \left( { \Pi_{02}}n_{12} \right) {{ \Pi_{01}}}^{4}
 -{\frac{28{{ m_2}}^{2}}{{{ m_1}}^{2}}} \left( { \Pi_{01}}{ \Pi_{02}} \right) {{
 \Pi_{01}}}^{4}
-{\frac {36 { m_2}}{{ m_1}}}  
  \left( { \Pi_{02}}n_{12} \right) ^{2}{{ \Pi_{01}}}^{4}
 \no & \quad
+{\frac {12 { m_2}}{{ m_1}}}\left( { \Pi_{01}}n_{12} \right) 
  ^{2}   \left( { \Pi_{02}}n_{12}  \right) ^{2
}{{ \Pi_{01}}}^{2}
+{\frac {8{ m_2}}{{m_1}}}  
 \left( { \Pi_{01}}{ \Pi_{02}} \right)^{2}{{ \Pi_{01}}}^{2}
\no & \quad
-12   \left( { \Pi_{02}}n_{12} \right) ^{
2} \left( { \Pi_{01}}{ \Pi_{02}} \right) {{ \Pi_{01}}}^{2}
-{
\frac {20 { m_1}}{{ m_2}}}   \left( { \Pi_{02}}n_{12}  
 \right) ^{4}{{ \Pi_{01}}}^{2}
 -4   \left( {
 \Pi_{02}}n_{12} \right)^{3} \left( { \Pi_{01}}n_{12} \right) 
{{ \Pi_{01}}}^{2}
\no & \quad
-17{{ \Pi_{02}}}^{2} \left( { \Pi_{01}}{ 
\Pi_{02}} \right) {{ \Pi_{01}}}^{2}
-{\frac{8{ m_2}}{{ m_1}}}{{ \Pi_{02}}}^{2}
 \left( { \Pi_{01}}n_{12} \right)^{2}{{ \Pi_{01}}}
^{2}
\nonumber\\&\quad
+{\frac {16 { m_2}}{{ m_1}}} \left( { \Pi_{01}}{
 \Pi_{02}} \right)  \left( { \Pi_{02}}n_{12} \right) 
 \left( { \Pi_{01}}n_{12} \right) {{ \Pi_{01}}}^{2}
+25{{ 
\Pi_{02}}}^{2} \left( { \Pi_{02}}n_{12} \right)  \left( { 
\Pi_{01}}n_{12} \right) {{ \Pi_{01}}}^{2}
\nonumber\\&\quad
-10 \left( { \Pi_{01}}n_{12}
 \right)  \left( { \Pi_{01}}{ \Pi_{02}}
 \right) ^{2} \left( { \Pi_{02}}n_{12} \right) -2  
 \left( { \Pi_{01}}{ \Pi_{02}} \right) ^{3}+5  
  \left( { \Pi_{01}}n_{12} \right)  ^{3}   \left( {
 \Pi_{02}}n_{12} \right) ^{3}
 \nonumber\\&\quad
 +15\left( { 
\Pi_{01}}n_{12}\right) ^{2} \left( { \Pi_{02}}n_{12}
 \right) ^{2} \left( { \Pi_{01}}{ \Pi_{02}} \right) 
\bigg]
\nonumber\\&+
{\frac {{G}^{2}}{144\,{r}^{2}}}
\bigg[
-{\frac {957 { m_2}{{ \Pi_{01}}}^{4}}{{{ m_1}}^{2}}}
-{\frac {261 {{ m_2}}^{2}{{ \Pi_{01}}}^{4}}{{{ m_1}}^{3}}}
-{\frac {90 {m_2}}{{{ m_1}}^{2}}} \left( { \Pi_{01}}{ \Pi_{02}} \right)
{{ \Pi_{01}}}^{2}
\no& \quad
+{\frac {654}{{ m_2}}}\left( { \Pi_{02}}n_{12}\right)^{2}
{{ \Pi_{01}}}^{2}
 +{\frac {798}{{ m_1}}}{{ 
\Pi_{02}}}^{2}{{ \Pi_{01}}}^{2}
+{\frac {1848 { m_2}}{{{ m_1}}^{2}}}
 \left({ \Pi_{01}}n_{12}\right) ^{2}{{ \Pi_{01}}}
^{2}
\no & \quad
-{\frac {705}{{ m_1}}} \left( { \Pi_{02}}
n_{12} \right)^{2}{{ \Pi_{01}}}^{2}
+{\frac {1938}{{ m_1}}}\left( { \Pi_{01}}{ \Pi_{02}} \right)
{{ \Pi_{01}}}^{2}
-{\frac {2310}{{ m_1}}} \left( { \Pi_{01}}n_{12} \right) 
 \left( { \Pi_{02}}n_{12} \right) {{ \Pi_{01}}}^{2}
\no & \quad
 +{\frac {36 {{ m_2}}^{2} }{{{ m_1}}^{3}}}
  \left( { \Pi_{01}}n_{12}  \right) ^{
2}{{ \Pi_{01}}}^{2}-{\frac {1428}{{ m_1}}}  
 \left( { \Pi_{01}}{ \Pi_{02}}  \right) ^{2}
 -{\frac { 3192}{{ m_1}}}  \left( { \Pi_{01}}n_{12} \right) ^{2}
 \left( { \Pi_{01}}{ \Pi_{02}} \right)
\no & \quad
 -{\frac { 1078}{{ m_1}}}  \left( { \Pi_{01}}n_{12} \right) ^{3}
 \left( { \Pi_{02}}n_{12} \right)
+{\frac {1146}{{ m_1}}} \left( { \Pi_{01}}n_{12} \right) ^{2}  
 \left( { \Pi_{02}}n_{12}\right) ^{2}
 \no & \quad
 +{\frac {3660}{{ m_1}}}
  \left( { \Pi_{01}}n_{12} \right)  \left( { \Pi_{01}}{ 
\Pi_{02}} \right)  \left( { \Pi_{02}}n_{12} \right) 
-{\frac {104{ m_2}}{{{ m_1}}^{2}}}\left( { \Pi_{01}}n_{12}  \right) ^{4}
\bigg]           
\nonumber\\&+
{\frac {{G}^{3}}{20160\,{r}^{3}}}
\bigg[
-501760{{ m_2}}^{2}{{ \Pi_{01}}}^{2}-{\frac {85680{{ m_2}}^{3}{
{ \Pi_{01}}}^{2}}{{ m_1}}}-496736{ m_1}{ m_2}{{ \Pi_{01}}}^{
2}
\no & \quad
-12915{{ m_2}}^{2}{\pi }^{2}{{ \Pi_{01}}}^{2}
+147840{ m_1}{
 m_2}{ \ln\left(\frac{r}{r_1}\right)}{{ \Pi_{01}}}^{2}+562256{{ m_1}}^{2}
 \left( { \Pi_{01}}{ \Pi_{02}} \right)
\no & \quad
+{\frac {30240{{ m_2}}^{3}}{{ m_1}}}\left( { \Pi_{01}}n_{12} \right) ^{2}
+174720{{ m_2}}^{2} \left( { \Pi_{01}}n_{12} \right) ^{2}
+977808{ m_1}{ m_2} \left( {\Pi_{01}}n_{12}  \right) ^{2}
\no & \quad
-982848{{ m_1}}^{2}\left( { \Pi_{01}}n_{12} \right)
  \left( { \Pi_{02}}n_{12} \right)
-38745{ m_1}{ m_2}{\pi }^{2} \left( { \Pi_{01}}n_{12}
 \right)  \left( { \Pi_{02}}n_{12} \right)
\no & \quad
 +38745{{ m_2}}^{2}{
\pi }^{2}   \left( { \Pi_{01}}n_{12} \right) ^{2}+
12915{ m_1}{ m_2}{\pi }^{2} \left( { \Pi_{01}}{ 
\Pi_{02}} \right) 
+547120{ m_1}{ m_2} \left( { \Pi_{01}}{ \Pi_{02}} \right) 
\no &  \quad
-149520{ m_1}{ m_2} \left( { 
\Pi_{01}}n_{12} \right)  \left( { \Pi_{02}}n_{12} \right) -147840{{ m_1
}}^{2}{ \ln\left(\frac{r}{r_1}\right)} \left( { \Pi_{01}}{ \Pi_{02}} \right) 
\nonumber\\ &  \quad
+443520{{ m_1}}^{2}{ \ln\left(\frac{r}{r_1}\right)} \left( { \Pi_{02}}n_{12}
 \right)  \left( { \Pi_{01}}n_{12} \right) -443520{ m_2}{ 
m_1}{ \ln\left(\frac{r}{r_1}\right)}   \left( { \Pi_{01}}n_{12} 
 \right) ^{2}
\bigg]
\nonumber\\&+
{\frac {1}{840}}{\frac {{G}^{4}}{{r}^{4}}}
\bigg[
315{{ m_1}}^{4}{ m_2}+17427{{ m_1}}^{3}{{ m_2}}^{2}-3080
{{ m_1}}^{3}{{ m_2}}^{2}\lambda
-6160{{ m_1}}^{3}{{ m_2}}^{2}
{ \ln\left(\frac{r}{r_1}\right)}
\bigg]
\nonumber\\&+1\longleftrightarrow 2.\nonumber 
 \end{align}        
We have again adapted the notation of Ref.\ \cite{ABF01} introduced after Eq.\
\eqref{e:7}.
Furthermore,
we have posed ${n}_{12}=\frac{{x_1}-{x_2}}{r}$.
The term ``$ +1\longleftrightarrow 2$'' represents the expression that precedes
but 
with interchanged particle indices, including the contributions that are already
symmetric in the particle indices,
particularly $n_{12}$ must be changed into $n_{21}=-n_{12}$ there. Note that it is
possible to remove the logarithm terms through a coordinate transformation
preserving the harmonicity conditions outside the bodies \cite{BF01b}.

The generators of infinitesimal spatial translations and rotations are given on the whole phase space
by Eqs.\ (\ref{e:53}, \ref{e:54}) as
\begin{eqnarray} \label{e:75}
P_i&=&\sum_{a=1}^2\Pi_{0ai}=\Pi_{01i}+\Pi_{02i},\\
\label{e:76}
J_i&=&\sum_{m=0}^{1}\sum_{a=1}^2\sum_{j,k=1}^3\epsilon_{ijk}x_{aj}^{(m)}\Pi_{m ak}\no
&=&\sum_{a=1}^2\sum_{j,k=1}^3 \epsilon_{ijk}x_{aj}\Pi_{0 ak}
+\sum_{a=1}^2\sum_{j,k=1}^3 \epsilon_{ijk}x_{aj}^{(1)}\Pi_{1 ak}.
\end{eqnarray}
For the dynamics,  only their restrictions to the constraint surface
are relevant as integrals of motion and
generators of symmetry transformations.
They can be computed by
eliminating $x^{(1)}$ and $\Pi_1$ from Eqs.\ (\ref{e:75}, \ref{e:76})
with the help of the constraints (\ref{e:chi}, \ref{e:omega}). The result is
\begin{align}
\label{e:P2pN}\left.P_i\right|_\Gamma=&\Pi_{01i}+\Pi_{02i},\\
\left. J_i \right|_\Gamma=&\sum_{a=1}^2\sum_{j,k=1}^3
\epsilon_{ijk}\bigg[
\bigg(
x_{aj}-\varepsilon^2 \frac{1}{m_a}\Phi_{1aj}(x,x^{(1)}(x,\Pi_0))
\bigg)\Pi_{0 ak}
\no&
-\varepsilon^3 \frac{1}{m_a}\Phi_{0aj}(x,x^{(1)}(x,\Pi_0))\Phi_{1ak}(x,x^{(1)}(x,\Pi_0))\bigg]
+\mathcal{O}\left(\varepsilon^4\right).
\label{e:J2pN}
\end{align}                                           
The angular momentum $\left. J_i \right|_\Gamma$ of the post-Newtonian
dynamics can be written explicitly as
\begin{align}
J_{3\text{pN}i}=&\,^0\!J_{\text{N}i}+\frac{1}{c^4}\,^2\!J_{\text{N}i}+
\frac{1}{c^6}\;\,^3\!J_{\text
  N i}, \label{e:80}\\[7pt] 
  \,^0\!J_{\text{N}i}=& \epsilon_{ijk}x_{1j} \Pi_{0 1k}+\epsilon_{ijk}x_{2j}
\Pi_{0 2k},
\nonumber\\[7pt]
\,^2\!J_{\text{N}i}=&-\epsilon_{ijk}
\frac{7G}{4{ m_1}} {{ \left( n_{12}{ \Pi_{01}} \right) { \Pi_{01j}}{
\Pi_{02k}}}} 
+\epsilon_{ijk} {\frac {G}{8{ m_1}{ m_2}\,r}}
\bigg[
\left( n_{12}{ \Pi_{02}} \right)^{2}{ m_1}-7{ m_1}{{ \Pi_{02}}}^{2}
\bigg] 
{x_{1j}}{ \Pi_{01}} _{{k}}
\nonumber\\&
-\epsilon_{ijk}{\frac {G}{8{m_1}{ m_2}\,r}}
\bigg[ \left( n_{12}{ \Pi_{01}}\right)^{2}{ m_2}-7{ m_2}{{
\Pi_{01}}}^{2} \bigg] { x_{1j}}{ \Pi_{02k}}
+1\longleftrightarrow 2,
\nonumber\\[7pt]
\,^3\!J_{\text Ni}=&
\epsilon_{{ijk}}{\frac {G}{24{{m_1}}^{2}{{ m_2}}^{2}}}
\biggl[ 
 9{ m_2} \left( n_{12}{ \Pi_{02}} \right)\left( n_{12}{ \Pi_{01}}\right) ^{2}
-3{ m_2}{{ \Pi_{01}}}^{2}{ } \left(n_{12}{ \Pi_{02}} \right)
-{\frac {3{{ m_2}}^{2}}{{ m_1}}}{{ \Pi_{01}}}^{2}{ }
   \left(n_{12}{ \Pi_{01}} \right)  
\nonumber\\ & \quad 
+6 {m_2} \left( n_{12}{ \Pi_{01}} \right) { } \left( { \Pi_{01}}{ \Pi_{02}}
\right)
+{\frac{10{{ m_2}}^{2}}{{m_1}}}
\left( n_{12}{ \Pi_{01}}\right)^{3} \bigg] {\Pi_{01j}}{\Pi_{02k}} 
\nonumber\\&
+ \epsilon_{{ijk}}
{\frac {G}{16{{ m_1}}^{2}{{ m_2}}^{2}r}} 
\bigg[ 7{ m_2}{{
\Pi_{02}}}^{2}{{ \Pi_{01} }}^{2}-{ m_2}{{ \Pi_{01}}}^{2} \left( n_{12}{
\Pi_{02}} \right) ^{2} +8{ m_1}{{ \Pi_{02}}}^{2} \left( n_{12} 
{ \Pi_{01}}\right) \left( n_{12}{\Pi_{02}} \right) 
\nonumber\\ & 
\quad -2{ m_1} { } \left( n_{12}{ \Pi_{01}} \right) \left( n_{12}{ \Pi_{02}}
\right) ^{3}
-{\frac {{{ m_1}}^{2}}{{ m_2}}} \left( n_{12}{ \Pi_{02}} \right) ^{4}-6{ m_1}
 \left( n_{12}{ \Pi_{02}} \right) ^{2}{ } \left( { \Pi_{01}}{ \Pi_{02}}
 \right) 
\nonumber\\ & \quad
+{\frac {14{{ m_1}}^{2}}{{ m_2}}}{{ \Pi_{02}}}^{4} 
+10{ m_2}{{ \Pi_{02}}}^{2} \left( n_{12}{ \Pi_{01}} \right)
 ^{2}+{\frac {3{{ m_1}}^{2 }}{{ m_2}}}{{ \Pi_{02}}}^{2} \left( n_{12}{ \Pi_{02}}
 \right) ^{2} \bigg]{ x_{1j}}{\Pi_{01k}} 
\nonumber\\&
  +\epsilon_{{ijk}} 
{\frac {G}{16{{ m_1}}^{2}{{ m_2}}^{2}r}}
 \bigg[
2{ m_2} \left( n_{12}{ \Pi_{02}} \right) \left( n_{12}{
 \Pi_{01}} \right) ^{3}
- {\frac {3{{ m_2}}^{2}}{{ m_1}}}{{ \Pi_{01}}}^{2} \left(
 n_{12}{ \Pi_{01}} \right) ^{2}
\no & \quad
-10{ m_1}{{ \Pi_{01}}} ^{2}
 \left(n_{12}{ \Pi_{02}}\right)^{2}  
-{\frac{14{{ m_2}}^{2}}{{ m_1}}}{{ \Pi_{01}}}^{4}
+{ m_1}{{ \Pi_{02} }}^{2}\left(n_{12}{ \Pi_{01}}\right)^{2}
\nonumber\\ & \quad
-8{ m_2}{{ \Pi_{01}}}^{2}\left(n_{12}{ \Pi_{02}} \right)
  \left( n_{12}{ \Pi_{01}} \right)
+{\frac {{{m_2}}^{2}}{{ m_1}}}\left(n_{12}{\Pi_{01}}\right)^{4}  
-7{ m_1}{{ \Pi_{02}}}^{2}{{ \Pi_{01}}}^{2}
\nonumber\\ & \quad
+6{ m_2} \left(
 n_{12} { \Pi_{01}} \right) ^{2} \left( { \Pi_{01}}{ \Pi_{02}} \right)
 \bigg] { x_{1j}}{ \Pi_{02k}} 
\nonumber\\ &
+\epsilon_{{ijk}}
{\frac {{G}^{2}}{24\,r}}
\bigg[
193 \left( n_{12}{ \Pi_{01}}\right) 
+{\frac{17{ m_2}}{{ m_1}}}\left(n_{12}{\Pi_{01}}\right)
 \bigg]
{\Pi_{01j}} { \Pi_{02k}}
\nonumber\\&+
\epsilon_{{ijk}}
{\frac {{G}^{2}}{48\,{r}^{2}}}
\bigg[
{\frac{21{ m_2}}{{ m_1}}} \left(n_{12}{ \Pi_{01}}\right)^{2}
-{\frac{408{ m_2}{{\Pi_{01}}}^{2}}{{ m_1}}}
-{\frac{68{ m_1}{{\Pi_{02}}}^{2}}{{ m_2}}}
+48\left( n_{12}{ \Pi_{01}} \right)\left( n_{12}{ \Pi_{02}} \right)
\nonumber\\ & \quad
 -109{{ \Pi_{02}}}^{2} +{\frac{100{ m_1}}{{ m_2}}}
 \left(n_{12}{ \Pi_{02}} \right) ^{2} -76\left( n_{12}{ \Pi_{02}}
 \right) ^{2}+ 816{ } \left( { \Pi_{01}}{ \Pi_{02}} \right) \bigg]{ x_{1j}}
 {\Pi_{01k}}
\nonumber\\&+ \epsilon_{{ijk}}{\frac
 {{G}^{2}}{48\,{r}^{2}}}
\biggl[
-{\frac{21{ m_1}}{{ m_2}}}
\left( n_{12}{ \Pi_{02}}\right)^{2}-{\frac{100{ m_2}}{{ m_1}}}
\left( n_{12}{ \Pi_{01}}\right)^{2} +76 \left( n_{12}{ \Pi_{01}} \right) ^{2}-816{ } \left( {
 \Pi_{01}}{ \Pi_{02}} \right) \biggr.\nonumber\\ & \biggl. \quad +{\frac {68{
 m_2}{{ \Pi_{01}}}^{2}}{{ m_1}}} +{ \frac {408{ m_1}{{ \Pi_{02}}}^{2}}{{
 m_2}}} +109{{ \Pi_{01}}}^{2}-48 { } \left( n_{12}{ \Pi_{01}} \right) \left(
 n_{12}{ \Pi_{02}} \right) \biggr] { x_{1j}}{ \Pi_{02k}} \nonumber\\
 &+
 1\longleftrightarrow 2,\nonumber
\end{align} 
where a sum over the indices $j$ and $k=1,2,3$ must be  understood. It has been checked that
the above momenta satisfy the relations (\ref{e:PA1}-\ref{e:PA5}).
In particular, $H$ is invariant under spatial translations and
rotations and the components of the generalized total
linear and angular momentum, $P_i$ and $J_i$, are integrals of
motion.

The generator \eqref{e:80} takes a simple familiar form when written in coordinates
that are canonically conjugate with respect to the Dirac bracket.
We have obtained these coordinates, expressed
in terms of $x, \Pi_0$ by guess work. They read
\begin{subequations}
\begin{eqnarray}
P_\alpha&=&\Pi_{0\alpha}-\varepsilon^3\sum_\gamma\frac{1}{m_\gamma}
\Phi_{1\gamma}(x,\Pi_0)\frac{\pa 
\Phi_{0\gamma}(x,\Pi_0)}{\pa
  x_\alpha}+\mathcal{O}\left(\varepsilon^{4}\right),
\label{e:82a}
\\
Q_\alpha&=&x_\alpha-\varepsilon^2\frac{1}{m_\alpha}\Phi_{1\alpha}(x,\Pi_0)
+\varepsilon^3\sum_{\gamma}\frac{1}{m_\gamma}
\Phi_{1\gamma}(x,\Pi_0) \frac{\pa \Phi_{0\gamma}(x,\Pi_0)}{\pa
\Pi_{0\alpha}}+\mathcal{O}\left(\varepsilon^{4}\right),
\label{e:82b}
\end{eqnarray}
\end{subequations}
where the sum $\sum_\gamma$ holds over all pairs of indices $\gamma=b\,j$, $b=1,2$,
$j=1,2,3$ and where $\Phi_{s\gamma}(x,\Pi_0)$, $s=0,1$, is a short notation for
$\Phi_{s\gamma}(x, x^{(1)}(x,\Pi_0))$. This result can be verified by
using the explicit expressions (\ref{e:EDB1}-\ref{e:EDB3})
for the elementary Dirac brackets.

It is important to stress that there does not exist any harmonic coordinate
system in which $Q$ and more generally any other canonical coordinates
represent the particle positions. If it were possible, we could go over to a
frame where the particle positions are canonical coordinates while maintaining
manifest-Poincar\'e invariance. In terms of these new coordinates, the
Hamiltonian as well as the Lagrangian would be ordinary although describing
manifest Poincar\'e invariant dynamics, in contradiction to the
no-interaction theorem \cite{MS79}. A consequence is that we are not able to obtain canonically
conjugate particle coordinates by means of Poincar\'e transformations. This can
be seen from Eqs.\ (\ref{e:C19}, \ref{e:C20}), which imply that
$\{x_\alpha,x_\beta\}^*$ cannot vanish everywhere for the 3pN and 2pC dynamics.
The form of these two relations is maintained by Poincar\'e transformations
that are also generalized
canonical transformations. However, in the case of vanishing
coupling constant, we have $Q\rightarrow x,P\rightarrow \Pi_0$. In other
words, for the non-interacting systems, manifest Poincar\'e invariance becomes
compatible with the choice of spatial coordinates as canonical coordinates.

The interest of using $Q$ and $P$ rather than $x$ and $\Pi_0$ for explicit
calculations
is that in these coordinates the Dirac bracket takes the simple
standard form of a Poisson bracket.
For simplicity, we shall note any function $F(x,\Pi_0)$
on the constraint surface expressed by means of the canonically conjugate
variables $Q,P$ as $\tilde F\equiv F(Q,P)\equiv F(x(Q,P),\Pi_0(Q,P))$.
The Dirac bracket at a point $x(Q,P),\Pi_0(Q,P)$
of the constraint surface then reduces to 
\begin{align}\label{e:DBQP}
\{F,G\}^*\bigg|_{x(Q,P),\Pi_0(Q,P)}
&
=\{\tilde F,\tilde G\}_{Q,P}
 \equiv\sum_{\alpha} \frac{\pa \tilde F}{\pa Q_\alpha}\frac{\pa \tilde G}{\pa
 P_\alpha}-\frac{\pa \tilde G}{\pa Q_\alpha} 
 \frac{\pa \tilde F}{\pa P_\alpha}.
\end{align}

In canonically conjugate coordinates, $P_i$ and
$J_i$ have 
their usual expression on the constraint surface
for both dynamics under consideration, namely
\begin{eqnarray}
\label{e:P3pN_kan}\tilde P_i&=&\sum_{a=1}^2 P_{ai}+\mathcal{O}\left(\varepsilon^{n+1}\right),
\\ \label{e:J3pN_kan}\tilde
J_i&=&\sum_{a=1}^2\sum_{j,k=1}^3\epsilon_{ijk}Q_{aj}P_{ak}
+\mathcal{O}\left(\varepsilon^{n+1}\right),
\end{eqnarray}
with $n=3$ at 3pN ($n=2$ at 2pC).
Indeed, spatial translations and rotations leave the constraint surface
invariant so that the 
restrictions of $P_i$ and $J_i$ to $\Gamma$
are the well known generators of infinitesimal
spatial translations and rotations on $\Gamma$
(cf. Sec.\  \ref{sec:5}) taking above shape
in terms of canonically conjugate coordinates.
This
explains incidentally the absence of a first post-Newtonian contribution in the expressions
\eqref{e:80} and \eqref{e:PJ2pC} for $\left.J_i\right|_\Gamma$ as a function of
$x,\Pi_0$. If the term containing $\Pi_{1}$ does not contribute to the first
order, this is because the $\Pi_{1}$'s are second order quantities; for the rest, the
$x$'s and the $\Pi_0$'s
are canonically conjugate modulo $1/c^4$ corrections.

To determine the generator of Lorentz boosts, it is
useful to work with the canonical coordinates $Q, P$ on the constraint
surface.
In addition, it is more convenient to
derive $\tilde K_i\equiv K_i(Q,P)=G_i(Q,P)+P_i(Q,P) t$ rather than $G_i(Q,P)$
itself.
We compute $\tilde K_i$ with the help of the
method of undetermined coefficients \cite{DJS00b}, thereafter
fixing the uniqueness of the solution.

As already mentioned, $K_i$ fulfills the same Dirac bracket relations
(\ref{e:PA6}-\ref{e:PA8}, \ref{e:PA9a})
as $G_i$, but it is not explicitly time dependent and it is not an integral of
motion.
The behavior of $\tilde K_i$ under spatial rotations is
governed by Eq.\ \eqref{e:PA8} with the angular momentum $J_i$ given by
\eqref{e:J3pN_kan}. We
conclude that $\tilde K_i$ has the general structure
\begin{equation}
\tilde K_i=\sum_{a=1}^2M_a(Q,P)Q_{ai}+N_a(Q,P) P_{ai}+\mathcal{O}\left(\varepsilon^{n+1}\right),
\end{equation}
where $M_a(Q,P)$, $N_a(Q,P)$ are two post-Newtonian scalar functions. The form
of the differential equations for $\tilde K_i$ resulting from the Poincar\'e algebra
and the world-line condition suggests for $M_a(Q,P)$ and $N_a(Q,P)$
the ansatz
\begin{align}
&c_{n_0,...,n_5}R^{n_0}P_1^{2n_1} P_2^{2n_2}(P_1P_2)^{n_3}
(N_{12}P_1)^{n_4}(N_{12}P_2)^{n_5}
\label{e:87}
+d_{1,s_1}R^{s_1} \ln(R/r_1)+d_{2,s_2}R^{s_2} \ln(R/r_2),
\end{align}
with $R=|Q_1-Q_2|$
and $N_{12}=\frac{Q_{1}-Q_{2}}{R}$;
the logarithm terms are only expected to appear in the function
$M_a$ and the powers
$n_0, s_1, s_2$ are presumed to be integers, while
$n_1,...,n_5$ are natural numbers. The admissible eight-tuples of $s_1, s_2,
n_0,...,n_5$ are restricted by demanding the correct  
physical dimension for $M_a$ and $N_a$.

Insertion of the preceding ansatz into the partial differential equations
(\ref{e:PA6}-\ref{e:PA8}, \ref{e:PA9a}) yields the linear system of equations
to be solved.
  The coefficients seem to be over-determined, but the equations
are not all independent so that there actually exists a solution. It is most easily
derived by means of a Computer-algebra
program\footnote{We have used the software Maple 8 \copyright  Waterloo Maple Inc.}.
If we only impose that the
Poincar\'e algebra should be satisfied, some coefficients remain
undetermined even at the 1pN order. They are set by requiring the world-line condition
on $K_i$, reflecting the manifest Lorentz invariance of the system. As a result,
the expression for $\tilde K_i$ obtained through this procedure,
is automatically consistent with both the Poincar\'e-algebra and the world-line condition.

Now, as discussed before, the canonically conjugate coordinates simplifying our
explicit calculation are not harmonic.
Re-expressing $\tilde K_i$ by the
coordinates $x$ and $\Pi_0$ of the constraint surface we finally arrive at
$K_i$ in a harmonic-coordinate frame. The
result is displayed up to third post-Newtonian order as
\begin{align}
K_{\text{3pN}i}=&\,^0\!K_{\text{N}i}+\frac{1}{c^2}\,^1\!K_{\text{N}i}+
\frac{1}{c^4}\,^2\!K_{\text{N}i}+\frac{1}{c^6}\,^3\!K_{\text{N}i},    \\[7pt]
\,^0\!K_{\text{N}i}=&m_1x_{1i}+m_2 x_{2i},\nonumber\\[7pt]
\,^1\!K_{\text{N}i}=&
 \frac{{\Pi_{01}}^2}{2m_1}x_{1i}
+\frac{{\Pi_{02}}^2}{2m_2}x_{2i}-\frac{G m_1 m_2}{2r}(x_{1i}+x_{2i}),
\nonumber\\[7pt]
\,^2K_{\text{N}i}=&-{\frac {{{\Pi_{01}}}^{4}}{8{{ m_1}}^{3}}}{ x_{1i}}
+
{\frac {G}{8{ m_1}{m_2}}}
 \bigg[
7{{m_2}}^{2}{{\Pi_{01}}}^{2}n_{12i}
-{{m_2}}^{2}\left( n_{12}{\Pi_{01}}\right)^{2}n_{12{i}}
-14{ m_1}{m_2}\left( n_{12}{\Pi_{02}} \right)\Pi_{01i}
\no & \quad
-14{{m_2}}^{2}\left( n_{12}{\Pi_{01}}\right)\Pi_{01i}
\bigg]
+{\frac {G}{4\,r}}
\bigg[
-{\frac {6{m_2}}{{ m_1}}}{{\Pi_{01}}}^{2}{ x_{1i}}
+ \left( n_{12}{\Pi_{01}} \right)\left(n_{12}{\Pi_{02}}\right){ x_{1i}}
\no & \quad
+7 \left( {\Pi_{01}}{\Pi_{02}}
 \right){ x_{1i}} 
\bigg]
+{\frac {{G}^{2}{ m_1}{m_2}}{4\,{r}^{2}}}
\bigg[
-5{ m_1}{ x_{1i}}
+7{m_2}{ x_{1i}}
\bigg]
+ 1 \longleftrightarrow 2,
\nonumber\\[7pt]
\,^3\!K_{\text{N}i}=&{\frac {{{ \Pi_{01}}}^{6}}{16{{ m_1}}^{5}}}x_{1i}
+
{\frac {G}{48{{ m_1}}^{2}{{ m_2}}^{2}}}
\bigg[
-{\frac {42{{ m_2}}^{3}}{{m_1}}}{{ \Pi_{01}}}^{4}n_{12i}
- {\frac {6{{ m_2}}^{3}}{{ m_1}}} \left( n_{12}{ \Pi_{01}} \right)
{{\Pi_{01}}}^{2}{ \Pi_{01i}} 
\no & \quad
-6{{ m_2}}^{2} \left( n_{12}{
\Pi_{02}} \right){{ \Pi_{01}}}^{2}{ \Pi_{01i}}
-30{ m_1}{m_2}\left( n_{12}{\Pi_{02}} \right)^{2}
{{ \Pi_{01}}}^{2}n_{12i}
\no & \quad
-24{{ m_2}}^{2}\left( n_{12}{\Pi_{02}}\right)
  \left( n_{12}{ \Pi_{01}}\right){{ \Pi_{01}}}^{2}n_{12i} 
-{\frac{9{{ m_2}}^{3}}{{ m_1}}}\left( n_{12}{ \Pi_{01}}\right)^{2}
  {{\Pi_{01}}}^{2}n_{12i}
\no & \quad  
+18{ m_1}{
m_2} \left( n_{12}{\Pi_{02}} \right) ^{2} \left( n_{12}{ \Pi_{01}} \right) {
  \Pi_{01i}} 
-48{ m_1}{ m_2}
{{\Pi_{02}}}^{2} \left( n_{12}{ \Pi_{01}} \right) { \Pi_{01i}}
\nonumber\\& \quad
+12{ m_1}{ m_2} \left( n_{12}{\Pi_{02}} \right)  \left( { \Pi_{01}}{\Pi_{02}}
\right) { \Pi_{01i}} 
+20{{ m_1}}^{2} \left( n_{12}{\Pi_{02}}\right) ^{3}{\Pi_{01i}}
\no & \quad
+{\frac{20{{ m_2}}^{3}}{{ m_1}}}
\left( n_{12}{\Pi_{01}} \right)^{3}{ \Pi_{01i}}
+18{{ m_2}}
^{2} \left( n_{12}{\Pi_{02}} \right)  \left( n_{12}{
 \Pi_{01}} \right)^{2}{ \Pi_{01i}}
\no & \quad
-48{{ m_1}}^{2}{{\Pi_{02}}}^{2}
    \left( n_{12}{\Pi_{02}} \right){ \Pi_{01i}}
+12{{ m_2}}^{2} \left( n_{12}{ \Pi_{01}} \right)
    \left({\Pi_{01}}{\Pi_{02}} \right) { \Pi_{01i}}
\nonumber\\& \quad
 +{\frac {3{{ m_2}}^{3}}{{ m_1}}} \left( n_{12}
{ \Pi_{01}}\right)^{4}n_{12i}+6{{ m_2}}^{2}
\left( n_{12}{\Pi_{02}} \right)
\left( n_{12}{ \Pi_{01}}\right) ^{3}n_{12i}
\no & \quad
 +18
{{ m_2}}^{2}n_{12i} \left( n_{12}{ \Pi_{01}}
 \right) ^{2} \left( { \Pi_{01}}{\Pi_{02}} \right)
\bigg]
\nonumber\\ & +
{\frac {G}{16\,r}}
\bigg[
{\frac{21{ m_2}}{{{ m_1}}^{3}}}{{ \Pi_{01}}}^{4}x_{1i}
+{\frac {14}{{ m_1}{ m_2}}} \left( n_{12}{\Pi_{02}}
 \right)^{2}{{ \Pi_{01}}}^{2}x_{1i}
+{\frac{2}{{{ m_1}}^{2}}}\left( { \Pi_{01}}{\Pi_{02}} \right) {{
 \Pi_{01}}}^{2}x_{1i}
\no & \quad 
 -{\frac{15}{{ m_1}{ m_2}}}{{\Pi_{02}}}^{2}{{ \Pi_{01}}}^{2} x_{1i}
-{\frac {2}{{{ m_1}}^{2}}} \left( n_{12}{ \Pi_{01}} \right)  \left( n_{12}{\Pi_{02}}
 \right) {{ \Pi_{01}}}^{2}x_{1i}
-{\frac {2}{{m_1}{ m_2}}}
 \left( { \Pi_{01}}{\Pi_{02}}\right) ^{2}x_{1i}
\no & \quad
-{\frac {12}{{ m_1}{ m_2}}} \left( n_{12}{\Pi_{01}}\right)\left( n_{12}{\Pi_{02}} \right)  \left( {
 \Pi_{01}}{\Pi_{02}} \right) x_{1i}
 +{\frac {2}{{{ m_2}}^{2}} {{\Pi_{02}}}^{2} \left( { \Pi_{01}}{\Pi_{02}} \right) }
 x_{1i}
\no & \quad
-{\frac {2}{{{m_2}}^{2}}}
 {{\Pi_{02}}}^{2}
 \left( n_{12}{\Pi_{02}} \right)  \left( n_{12}{ \Pi_{01}} \right)
 x_{1i}
-{\frac {3}{{ m_1}{ m_2}}}\left( n_{12}{
 \Pi_{01}} \right) ^{2} \left( n_{12}{\Pi_{02}}
 \right)^{2}x_{1i} 
\no & \quad
 -{\frac {11{ m_1}}{{{ m_2}}^{3}}}{{\Pi_{02}}}^{4}x_{1i}
\bigg]
\nonumber\\&+
{\frac {{G}^{2}}{48\,r}}
\bigg[
-299{ m_2}{{ \Pi_{01}}}^{2}n_{12i}+{\frac{68{{ m_2}}^{2}}{{ m_1}}}
{{ \Pi_{01}}}^{2}n_{{12i}}
+{\frac{34{{ m_2}}^{2}}{{ m_1}}}\left( n_{12}{ \Pi_{01}} \right) { \Pi_{01i}}
\no & \quad
+428{ m_2}\left( n_{12}{ \Pi_{01}} \right) { \Pi_{01i}}
+386{ m_2}
 \left( n_{12}{\Pi_{02}} \right) { \Pi_{01i}}+76{ m_1}
 \left( n_{12}{\Pi_{02}} \right) { \Pi_{01i}}
\no & \quad
+48{ m_1} \left(
 n_{12}{\Pi_{02}} \right)  \left( n_{12}{ \Pi_{01}} 
 \right) n_{12i}
 -{\frac {100{{ m_2}}^{2}}{{ m_1}}}\left(
  n_{12}{ \Pi_{01}} \right)^{2}n_{12i}
  +13{ m_2}
   \left( n_{12}{ \Pi_{01}}  \right) ^{2}n_{12i}
\no & \quad
   +816
{ m_1} \left( { \Pi_{01}}{\Pi_{02}} \right)n_{12i}
 \bigg]
\nonumber\\&+
{\frac {1}{3360}}{\frac {{G}^{2}}{{r}^{2}}}
\bigg[
21140{ m_2} \left( { \Pi_{01}}{\Pi_{02}} \right) x_{1i}-50540{ m_1} \left( {
 \Pi_{01}}{\Pi_{02}} 
 \right) x_{1i}
  +21210{ m_2}{{ \Pi_{01}}}^{2}x_{1i}
\nonumber\\& \quad
  -10640{ m_1} \left( n_{12}{ \Pi_{01}} \right)
 \left( n_{12}{\Pi_{02}} \right) x_{1i}
 -1470{ m_1}{{\Pi_{02}
}}^{2}x_{1i}
 -5670{ m_2} \left( n_{12}{
\Pi_{01}} \right)^{2}x_{1i}
 \nonumber\\& \quad
-{\frac {1680{{ m_2}}^{2}}{{ m_1}}}
 \left( n_{12}{ \Pi_{01}}\right) ^{2}x_{1i}
 +{\frac{9240{{ m_2}}^{2}}{{ m_1}}}{{ \Pi_{01}}}^{2}
 x_{1i}
  +17360{ m_2} \left( n_{12}{ \Pi_{01}} \right)
 \left( n_{12}{\Pi_{02}} \right) x_{1i}
 \nonumber\\& \quad
 +210{ m_1}
 \left( n_{12}{\Pi_{02}} \right) ^{2}x_{1i}
 \bigg]
\nonumber\\&+
{\frac {1}{2520}}{\frac {{G}^{3}}{{r}^{3}}}
\bigg[
-5985{{ m_1}}^{2}{{ m_2}}^{2}x_{1i}
-28702{ m_1}{{m_2}}^{3}x_{1i}
-18480{{ m_1}}^{3}{ m_2}{\ln\left(\frac{r}{r_1}\right)}x_{1i}
\no & \quad
+27442{{ m_1}}^{3}{ m_2}x_{1i} 
+18480{ m_1}{{ m_2}}^{3}{ \ln\left(\frac{r}{r_2}\right)}x_{1i}
\bigg]
\no &
+ 1 \longleftrightarrow 2.\nonumber
\end{align}

The 3pN accurate generators $H_{3\text{pN}},
P_{3\text{pN}i}, J_{3\text{pN}i},
G_{3\text{pN}i}=K_{3\text{pN}i}-P_{3\text{pN}i}t$
we have derived 
in harmonic coordinates,
constitute an approximate representation of the Poincar\'e Algebra given
by the generators
of infinitesimal canonical transformations
with respect to the Dirac bracket, and thus,
they generate the Poincar\'e transformation group on the
phase space coordinates. The quantities are
integrals of motion that may be interpreted physically as the
3pN conserved total energy, the generalized total linear momentum,
the generalized total angular momentum and the center of mass constant.
Their expressions have been checked up to first order by comparing them with
Ref.\ \cite{SH76}.
If we adapt the ``Lagrangian-like'' coordinates $x, x^{(1)}$ on the constraint surface,
the above 
generators of
infinitesimal canonical transformations representing conserved quantities
reproduce the Noetherian
constants of motion associated with the Poincar\'e symmetry of the dynamics
that are
derived in \cite{ABF01} by means of the Lagrangian formalism. Note
that the definitions of $K_i$ and $G_i$ in this article are reverse.

It remains to show the uniqueness of the generators 
$P_i,J_i,G_i$ 
of the Poincar\'e group up to generalized canonical transformations provided
that $H$ is a
post-Newtonian Hamiltonian.
Let us indicate that the following proof is also valid for the post-Coulombian
dynamics.

 We start with the case of the momenta $P_i$ and $J_i$. Their uniqueness as a
 function of the phase space coordinates has been established 
on the whole phase space in Sec.\ \eqref{sec:4}. It entails
the uniqueness of their restriction to the constraint surface.

We next consider the center of mass constant $G_i$. 
Its uniqueness 
is equivalent to that
of $\tilde K_i\equiv K_i(Q,P)=G_i(Q,P)+P_i(Q,P)t$, since the coordinate
transformation is a diffeomorphism and since $P_i$ is unique. Again, $Q,P$ denotes
the set of canonically conjugate coordinates on the constraint surface
while $P_i$ represents the total linear momentum of the system.
Let us assume that
there exist two solutions $\tilde K_i$ and
$\tilde K_i'=\tilde K_i+f_i(Q,P)$ that both approximately 
fulfill the Poincar\'e algebra and the world-line condition
within the common coordinate frame.
The relations (\ref{e:WB}, \ref{e:PA7}, \ref{e:PA8}) written
in terms of the coordinates $Q,P$ as
given by Eqs.\ (\ref{e:82a}, \ref{e:82b}) read (cf. also \eqref{e:DBQP}):
\begin{align}
\{\tilde x_{ai},\tilde K_j\}_{Q,P} = &
\bigg\{Q_{ai}+\varepsilon^2\frac{1}{m_a}\tilde
  \Phi_{1ai}-\varepsilon^3\sum_{\gamma}\frac{1}{m_\gamma}\tilde \Phi_{1\gamma} 
\frac{\pa \tilde \Phi_{0\gamma}}{\pa P_{ai}},
\,^0\!\tilde K_j+\varepsilon\, \,^1\!\tilde K_j+
\varepsilon^2\, \,^2\!\tilde K_j
+\varepsilon^3\, \,^3\!\tilde K_j\bigg\}_{Q,P}
\no 
&+\mathcal{O}\left(\varepsilon^4\right)
\label{e:EBW1}
=\varepsilon \tilde x_{aj} \{\tilde x_{ai},\tilde H\}_{Q,P}
+\mathcal{O}\left(\varepsilon^4\right),\\
\{\tilde K_j,\tilde H\}_{Q,P}  = & 
\tilde{P}_j
+\mathcal{O}\left(\varepsilon^4\right),\label{e:EBW2}\\
\label{e:EBW3}
\{\tilde J_i, \tilde K_j\}_{Q,P}=&\sum_{k=1}^3\epsilon_{ijk}\tilde K_k
+\mathcal{O}\left(\varepsilon^4\right).
\end{align}
Note that,because $x_\alpha, \Pi_{0\beta}$ and $Q_\alpha, P_\beta$ agree at
zeroth order, the Newtonian (or Coulombian) contribution to the Hamiltonian in canonically
conjugate coordinates has the form
\begin{equation}\label{e:HEBW}
\,^0\!\tilde H=\sum_{a=1}^2 \frac{P_a^2}{2m_a}+U(Q).
\end{equation}
The uniqueness of $\tilde K_i$ is then proved order by order
using Eqs.\ (\ref{e:EBW1}-\ref{e:HEBW}).
First, we may insert successively
$\tilde K_j$ and $\tilde K_j'$ into \eqref{e:EBW1} since they are both
assumed to fulfill the latter relation. Taking the zeroth
order of the difference, we find that
\begin{equation}
\frac{\pa \,^0\!f_j(Q,P)}{\pa P_{ai}}=0,
\end{equation}
hence  $\,^0\!f_j(Q,P)=\,^0\!f_j(Q)$.
Next, we go over to Eq.\ \eqref{e:EBW2},
where we insert again $\tilde K_j$ and $\tilde K_j'$ successively before
subtracting the ensuant equalities. By virtue of Eq.\ \eqref{e:HEBW} we obtain at
zeroth order:
\begin{equation}
\label{e:93}
\sum_{a=1}^2\sum_{k=1}^3\frac{P_{ak}}{m_a}\frac{\pa \,^0\!f_j(Q)}{\pa
  Q_{ak}}=0. 
\end{equation}
Because $Q_\alpha, P_\beta$ constitute a set of independent coordinates,
and Eq.\ \eqref{e:93} holds for all $P_{a k}$, we
are led to
\begin{equation}
\frac{\pa \,^0\!f_j(Q)}{\pa Q_{ak}}=0  \quad \Rightarrow \quad \,^0\!f_j(Q)=\text{const}.
\end{equation} 
From  Eq.\ \eqref{e:EBW3} taken at zeroth order, we conclude by applying the same
procedure as explained above, that 
\begin{align} 
0&=\{\,^0\!\tilde{J}_i,  \,^0\!f_j\}_{Q,P}=\{\,^0\!\tilde{J}_i, \,^0\!\tilde
K_j'-\,^0\!\tilde K_j\}_{Q,P}\nonumber\\ 
&=\sum_{k=1}^3\epsilon_{ijk}(\,^0\!\tilde K_k'-\,^0\!\tilde K_k)=
\sum_{k=1}^3\epsilon_{ijk}\,^0\!f_k \nonumber \\
\label{e:Nach2}
&\Rightarrow \quad \,^0\!f_k=0 \quad \Rightarrow \quad \,^0\! \tilde
K_i'=\,^0\!\tilde K_i.
\end{align}
The proof of uniqueness at the first order is similar. The world-line condition
Eq.\ \eqref{e:EBW1}, truncated at this level of approximation yields
$\,^1\!f_j(Q,P)=\,^1\!f_j(Q)$. Eq.\ \eqref{e:EBW2} reduces to
\begin{equation}\label{e:Nach3}
\{\,^1\!\tilde K_j, \,^0\!\tilde H\}_{Q,P}+\{\,^0\!\tilde K_j, \,^1\!\tilde
H\}_{Q,P}=0,
\end{equation}
and we have the same equation for $\tilde K_j'$. On the other hand,
we know from Eq.\ \eqref{e:Nach2}
that $\tilde K_{i}'$ and $\tilde K_{i}$ differ at most
from the first order on. Therefore taking the difference between the equation
\eqref{e:Nach3} for $\tilde K_j'$ and the same equation for
$\tilde K_j$ at first order leads to
\begin{eqnarray}
 \{\,^1\!f_j(Q), \,^0\!\tilde H\}_{Q,P}&=&0\nonumber\\
 \sum_{a=1}^2\sum_{k=1}^3\frac{P_{ak}}{m_a}\frac{\pa \,^1\!f_j(Q)}{\pa
 Q_{ak}}&=&0\nonumber\\
\label{e:Nach4} \Rightarrow \quad \frac{\pa \,^1\!f_j(Q)}{\pa
 Q_{ak}}&=&0,
\end{eqnarray}
so that $\,^1\!f_j(Q)=\text{const}$. Now, from the first order truncated
version of Eq.\ \eqref{e:EBW3}, it follows
\begin{align} 
0&=\{\,^0\!\tilde{J}_i,  \,^1\!f_j\}_{Q,P}=\{\,^0\!\tilde{J}_i, \,^1\!\tilde
K_j'-\,^1\!\tilde K_j\}_{Q,P}\nonumber\\ 
&=\sum_{k=1}^3\epsilon_{ijk}(\,^1\!\tilde K_k'-\,^1\!\tilde K_k)=
\sum_{k=1}^3\epsilon_{ijk}\,^1\!f_k\nonumber\\
\label{e:Nach5}
&\Rightarrow \quad \,^1\!f_k=0 \quad \Rightarrow \quad \,^1\! \tilde
K_i'=\,^1\!\tilde K_i. 
\end{align}                                               
The uniqueness of second, third or even higher orders
follows analogously.

\section{\label{sec:6}Application to $2$pC dynamics} 

The Lagrangian describing the binary
dynamics of the Feynman-Wheeler theory in Lorentz gauge up to second
post-Coulombian order takes the form
\begin{align}
L(x, x^{(1)}, x^{(2)}) =& \frac{1}{2}\sum_{a=1}^2 m_a
(x_a^{(1)})^2+V_0(x) +
\frac{1}{c^2}V_1(x,x^{(1)}) 
+
\frac{1}{c^4}V_2(x,x^{(1)},x^{(2)}),
\end{align}
with $V_0(x)=-\f{e_1 e_2}{r}$.
We may thus adopt directly all general results derived for the $3$pN dynamics by
neglecting the third order contribution.
The elementary Dirac brackets of the 2pC dynamics, for instance, may be inferred from
appendix \ref{sec:C}.
The computation of the
generators of the Poincar\'e group corresponding to the
conserved quantities of the Feynman-Wheeler 2pC binary dynamics
can be performed in a way similar
to the post-Newtonian case. The results read as follows.

 The reduced Hamiltonian is given by
\begin{align}\label{e:2pCH}
H_{2\text{pC}}=&\,^0\!H_\text{C}+\frac{1}{c^2}\,^1\!H_\text{C}+\frac{1}{c^4}
\,^2\!H_\text{C},\\[7pt]
\,^0\!H_\text{C}=&\f{{\Pi_{01}}^2}{2m_1}+\f{{\Pi_{02}}^2}{2m_2}+\f{e_1
e_2}{r},
\nonumber\\[7pt]
\;^1\!H_\text{C}=&
-{\frac {{{\Pi_{01}}}^{4}}{8{{m_1}}^{3}}}
-{\frac {{{\Pi_{02}}}^{4}}{8{{m_2}}^{3}}}
-{\frac {{ e_1}{ e_2}}{2{m_1}{m_2}\,r}}
\bigg[ 
\left({\Pi_{01}}{\Pi_{02}}\right) + \left(n_{12}{\Pi_{02}}\right)
\left(n_{12}{\Pi_{01}}\right)
\bigg],
\nonumber\\[7pt]
 \,^2H_\text{C}= &
{\frac {{{\Pi_{01}}}^{6}}{16{{ m_1}}^{5}}} + 
{\frac {{ e_1}{ e_2}}{16{{m_1}}^{2}{{ m_2}}^{2}\,r}}
\bigg[
 {{\Pi_{02}}}^{2}{{\Pi_{01}}}^{2}
-2 \left(
  {\Pi_{02}} n_{12} \right)^{2}{{ \Pi_{01}}}^{2}
+{\frac{4{ m_2}}{{m_1}}} \left({\Pi_{02}}n_{12} \right) \left( { \Pi_{01}}n_{12}
	\right){{ \Pi_{01}}}^{2} 
\nonumber\\&\quad
+{\frac{4{ m_2}}{{ m_1}}}\left( { \Pi_{01}}{\Pi_{02}} \right)
{{ \Pi_{01}}}^{2}
  -2 \left( { \Pi_{01}}{\Pi_{02}} \right) ^{2} + 3 \left( {
   \Pi_{01}}n_{12} \right) ^{2} \left( {\Pi_{02}}n_{12}\right) ^{2} \bigg]
   \nonumber\\&+{\frac {{{ e_1}}^{2}{{ e_2}}^{2}}{8{ m_1}^2{m_2}\,{r}^{2}}} 
\bigg[ {{ \Pi_{01}}}^{2}+3 \left( { \Pi_{01}}n_{12} \right)
     ^{2} \bigg] + {\frac {{{ e_1}}^{3}{{ e_2}}^{3}}{8{ m_1}{ m_2}\,{r}^{3}}}
   +1\longleftrightarrow 2.\nonumber
\end{align}
This result confirms the one derived in Ref.\ \cite{JLM86} as corrected in
Ref.\ \cite{DS88}.

The generators
of spatial translations and rotations $P_i$ and $J_i$ can be derived from Eqs.\ \eqref{e:P2pN}
and \eqref{e:J2pN}. The explicit expressions for the restrictions on the constraint
surface in terms of $x$ and $\Pi_0$ read
\begin{align}
\label{e:PJ2pC}
P_{\text{2pC}i}=&\Pi_{01i}+\Pi_{02i},\\[7pt]
J_{\text{2pC}i}=&\,^0\!J_{\text{C}i}+\frac{1}{c^4}\,^2\!J_{\text{C}i},\\[7pt]
 \,^0\!J_{\text{C}i}=& \epsilon_{ijk}x_{1j} \Pi_{0 1k}+\epsilon_{ijk}x_{2j}\Pi_{0 2k},\no[7pt]
 \,^2\!J_{\text{C}i}=&\epsilon_{ijk}
 \frac{{e_1}{e_2}}{4{ m_2}{{ m_1}}^{2}} {\left( { \Pi_{01}}n_{12} \right) {\Pi_{01j}} { \Pi_{02k }
 }}
+ \epsilon_{ijk}{\frac{{e_1}{e_2}}{8{{m_1}}^{2}{{m_2}}^{2}\,r}}
\bigg[ { m_1}{{ \Pi_{02}}}^{2}
-{ m_1} \left({\Pi_{02}}n_{12} \right) ^{2} \bigg] { x_{1j}}{\Pi_{01k}}
\nonumber \\&
+\epsilon_{ijk}{\frac {{ e_1}{ e_2}}{8{{ m_1}}^{2}{{ m_2}}^{2}\,r}}
 \bigg[ 
-{ m_2}{{ \Pi_{01}}}^{2} 
+ { m_2} \left( { \Pi_{01}}n_{12} \right)^{2} \bigg]{ x_{1j}} { \Pi_{02k}} 
\no &
- \epsilon_{ijk}{\frac {{{
 e_1}}^{2}{{ e_2}}^{2}}{4{ m_1}{ m_2}\,{r}^{2}} }\bigg[{
 x_{1j}}{\Pi_{01k}}-x_{1j}{ \Pi_{02k}}\bigg]
 +1\longleftrightarrow 2,\nonumber
\end{align}
where a sum over the indices $j$ and $k=1,2,3$ must be understood.

 The generator $G_i$ is computed as in the
post-Newtonian case. For $K_i=G_i+P_i t$ in terms of
$x$ and $\Pi_0$, we have
\begin{align}
K_{\text{2pC}i}=&\,^0\!K_{\text{C}i}+\frac{1}{c^2}\,^1\!K_{\text{C}i}+
\frac{1}{c^4}\,^2\!K_{\text{C}i},\\[7pt] 
\,^0\!K_{\text{C}i}=&m_1x_{1i}+m_2 x_{2i},\no[7pt]
\,^1\!K_{\text{C}i}=&
 \frac{{\Pi_1}^2}{2m_1}x_{1i}
+\frac{{\Pi_2}^2}{2m_2}x_{2i}
+\frac{e_1e_2}{2\,r}[x_{1i}+x_{2i}],
\no[7pt]
\,^2\!K_{\text Ci}=&-
{\frac {{ x_{1i}}{{\Pi_{01}}}^{4}}{8{{ m_1}}^{3}}}
+{\frac {{ e_1}{ e_2}}{8{ m_1}{ m_2}}}
\bigg[
 {\frac {{ m_2}}{{ m_1}}}\left( n_{12} {\Pi_{01}} \right) ^{2}n_{{12i}} 
-{\frac {{ m_2}}{{ m_1}}}{{ \Pi_{01}}}^{2}n_{{12i}}
+2\left( n_{12}{ \Pi_{02}} \right) { \Pi_{01i}}
\nonumber \\&
 +{\frac {2{ m_2}}{{ m_1}}} \left( n_{12}{ \Pi_{01}} \right) { \Pi_{01i}}
  \bigg]
- {\frac {{ e_1}{ e_2}}{4{ m_1}{ m_2}\,r}}
  \bigg[ \left( { \Pi_{01}}{ \Pi_{02}} \right){ x_{1i}}+\left( n_{12}{ \Pi_{01}}
  \right) \left( n_{12}{ \Pi_{02} } \right){ x_{1i}} \bigg]
\no &
-{\frac {{{ e_1}}^{2}{{ e_2}}^{2}}{4{ m_2}\,r}}n_{{12i}}
+ 1\longleftrightarrow 2.\nonumber
\end{align}
The above generators $H, P_i, J_i$ and $G_i$ of infinitesimal generalized canonical
transformations provide a representation of the Poincar\'e algebra on the constraint
surface of 2pC Feynman-Wheeler binary dynamics with respect
to the Dirac bracket.
The uniqueness of $P_i, J_i$ and $G_i$ up to a generalized canonical
transformation
has been established in Sec.\ \ref{sec:4}.

\section{\label{sec:7} Summary and Discussion }
In the present article, we applied an appropriate canonical formalism to
describe the third post-Newtonian dynamics of point-mass binaries in harmonic
coordinates. We treated the second order post-Coulombian dynamics of
Feynman-Wheeler theory in Lorentz gauge analogously.
In contrast to earlier works, we did not leave the coordinate
conditions by performing a higher order
contact transformation \cite{DS91} but instead we generalized a method
developed in Ref.\
\cite{JLM86} and constructed the dynamics directly in harmonic coordinates
resp. Lorentz gauge within the
framework of canonical formalism both singular and of higher order in the time
derivatives.  The canonical formulation
opens the way to advanced investigations about the geometrical and
physical interpretation of the motion.
It is highly desirable for deriving the
generators of infinitesimal generalized canonical symmetry transformations
that provide the integrals of motion.
We computed for the first time the generators of the Poincar\'e
 transformation group or,
equivalently, the conserved quantities corresponding to the manifest
Poincar\'e invariance, for third post-Newtonian conservative binary
dynamics in harmonic coordinates
as well as for Feynman-Wheeler second post-Coulombian
binary dynamics in Lorentz gauge.
An appropriate choice of  coordinates of the constraint surface reveals that
the 3pN conserved quantities we have obtained
agree with those
derived via the generalized Noether theorem in
\cite{ABF01}. After being reduced to the
center of mass frame \cite{BI03}, they can be used for the derivation of an
analytic parametric solution to the third post-Newtonian equations of motion
in harmonic gauge for compact binaries in eccentric orbits \cite{MGS04}, which
is in turn of high practical relevance for the construction of gravitational
wave search templates and comparison with
numerical simulations.
A useful generalization of the present investigations
is suggested to be the application of our
method to post-Newtonian binary dynamics in harmonic coordinates including
spin.

\begin{acknowledgments}
The authors are thankful to Guillaume Faye for a careful reading of the manuscript
and many useful comments. The financial support of the Deutsche Forschungsgemeinschaft (DFG)
through SFB/TR7 ``Gravitationswellenastronomie'' is gratefully acknowledged.
\end{acknowledgments}

\appendix

\section{\label{sec:A}Lagrangian constraints \cite{JLM86}}

In this appendix
we display for completeness the iterative method developed in Ref.\ \cite{JLM86}
to derive a minimal set
of time stable constraints for a dynamics described by a Lagrangian
with the structure shown in Eq.\ \eqref{e:8}.

(i) The first stage of the method consists in transforming the primary constraints
into a more restrictive set
implied by the primary constraints through time stability, without making use of
the time evolution equations.
In this new set, the higher order
time derivatives are found separately in the various equations.

The primary constraints \eqref{e:13}
 imply
\begin{equation}\label{e:14}
\varepsilon^n
x_{\alpha}^{(2)}=\varepsilon^n\frac{1}{m_\alpha}A_{\alpha0}(x)+
\mathcal{O}(\varepsilon^{n+1}).
\end{equation}                                                       
Thus, Eqs.\ \eqref{e:13} multiplied by $\varepsilon^{n-1}$ are fulfilled by the
motion, as well as the relations following from \eqref{e:14} through repeated
time differentiation
\begin{equation}\label{e:15}
\varepsilon^n
x_{\alpha}^{(2+r)}=\varepsilon^n\frac{1}{m_\alpha}B_{\alpha,2+r,0}(x,x^{(1)})+
\mathcal{O}(\varepsilon^{n + 1} ),
\end{equation}
where $r=0,...,2n-3$; The quantities
$B_{\alpha,2+r,0}(x,x^{(1)})$ are derived from the $A_{\alpha0}$'s by differentiating
$r$ times and eliminating accelerations by virtue of Eq.\ \eqref{e:14} whenever they occur;
the relation \eqref{e:15} reduces to \eqref{e:14} when $r=0$. Beware that,
 because of the occurrence of the time derivative
$x_{\alpha}^{(2n)}$, Eq.\ \eqref{e:15} with $r=2n-2$ is not a constraint, but rather
the condition of time stability for the actual constraints given by the system
\eqref{e:15}
for $r=0,...,2n-3$. It becomes a consequence of the
equations of motion as soon as the dynamics is restricted to the constraint
surface.

The next set of constraints implied by the primary ones
can be found by multiplying Eqs.\ \eqref{e:13} by $\varepsilon^{n-2}$:
\begin{equation}\label{e:16}
\varepsilon^{n-1}
x_{\alpha}^{(2)}=\varepsilon^{n-1}\frac{1}{m_\alpha}\left[A_{\alpha 0}(x)+
\varepsilon
A_{\alpha 1}(x,x^{(1)},x^{(2)})\right]+\mathcal{O}(\varepsilon^{n+1}).
\end{equation}
The acceleration dependence of $A_{\alpha1}(x,x^{(1)},x^{(2)})$ may be
eliminated by means of Eqs.\ \eqref{e:15}
\begin{equation}\label{e:17}
\varepsilon^{n-1}
x_{\alpha}^{(2)}=\varepsilon^{n-1}\frac{1}{m_\alpha}\left[A_{\alpha0}(x)+
\varepsilon
B_{\alpha,2,1}(x,x^{(1)})\right]+\mathcal{O}(\varepsilon^{n+1}).
\end{equation}
Differentiating $r=0,...,2n-3$ times with respect to time
and replacing all occurring accelerations by means of Eqs.\ \eqref{e:17}
results in
\begin{equation}\label{e:18}
\varepsilon^{n-1}
x_{\alpha}^{(2+r)}=\varepsilon^{n-1}
\frac{1}{m_\alpha}\left[B_{\alpha,2+r,0}(x,x^{(1)})
+\varepsilon B_{\alpha,2+r,1}(x,x^{(1)})\right]+\mathcal{O}(\varepsilon^{n+1}).
\end{equation}

Carrying on with this procedure we arrive at the set of constraints
\begin{equation}\label{e:19}
\varepsilon
x_{\alpha}^{(2+r)}=\varepsilon\frac{1}{m_\alpha}\left[\sum_{s=0}^{n-1}
\varepsilon^s B_{\alpha,2+r,s}(x,x^{(1)})\right]
+\mathcal{O}(\varepsilon^{n+1}),
\end{equation}             
where $r=0,...,2n-3$.

(ii) In the second part of the method proposed by Jaen, Llosa and Molina, the above system
is completed to one that is stable under the time evolution
ruled by the Euler-Lagrange equations.
The time stability condition for Eqs.\ \eqref{e:19}
which also covers the time stability of all previous sets of constraints,
equals \eqref{e:19} with $r=2n-2$:
\begin{equation}\label{e:20}
\varepsilon
x_{\alpha}^{(2n)}=\varepsilon\frac{1}{m_\alpha}
\left[\sum_{s=0}^{n-1}\varepsilon^sB_{\alpha,2n,s}(x,x^{(1)})\right]
+\mathcal{O}(\varepsilon^{n+1}).
\end{equation}
Demanding time stability of the constraints for the motion on the constraint
surface is equivalent to removing higher order time derivatives in the
Euler-Lagrange equations with the help of Eqs.\ \eqref{e:19} and eliminating
$x_{\alpha}^{(2n)}$ by means of Eq.\ \eqref{e:20}. We obtain the new set of
constraints
\begin{equation}\label{e:21}
x_{\alpha}^{(2+r)}=\frac{1}{m_\alpha}\left[\sum_{s=0}^{n}\varepsilon^s
  B_{\alpha,2+r,s}(x,x^{(1)})\right] 
+\mathcal{O}(\varepsilon^{n+1}),
\end{equation}
with  $r=0,...,2n-3$.
From the way this set of constraints has been obtained, we see that it must
hold even if some of the $ x_{\alpha}^{(2n)}$ do not occur in certain linear combinations of
the Euler-Lagrange equations, which may be the case for non-invertible
$\frac{\pa^2 V_n}{\pa x_{ai}^{(n)}\pa x_{bj}^{(n)}}$.

For the set of constraints \eqref{e:21},
 the time
stability condition reads
\begin{equation}\label{e:22}
x_{\alpha}^{(2n)}=\frac{1}{m_\alpha}\left[\sum_{s=0}^{n}
\varepsilon^s B_{\alpha,2n,s}(x,x^{(1)})\right]
+\mathcal{O}(\varepsilon^{n+1}).
\end{equation}
This condition is satisfied in the sense that removing the higher derivatives
in the Euler-Lagrange equations with the help of Eqs.\ \eqref{e:21} and substituting
$x_{\alpha}^{(2n)}$ by Eq.\ \eqref{e:22} yields a constraint (Eq.\
\eqref{e:21} with $r=0$) fulfilled on the constraint surface. It
is obvious, that the set \eqref{e:21} is more restrictive than
the original primary constraints. Nonetheless, we see from the derivation that
it follows from them through the time stability condition.
As a result,
with \eqref{e:21}, we have found a minimal set of constraints stable in time
for the dynamics under investigation.

\section{\label{sec:B}3pN constraints}

In this appendix we derive the constraints for the two-body system
at the third post Newtonian order by modifying appropriately the method displayed
in appendix \ref{sec:A}.
  
Within our approximation scheme, any vector being a multiple of
$\varepsilon^2$ is a null-vector of the Hessian of the 3pN Lagrangian
\eqref{e:L3pN}. Therefore, using Eq.\ 
\eqref{e:3pNeqm}
 as initial primary constraints, we have
\begin{equation}\label{e:pZB0a}
 \varepsilon^2\left(m_\alpha x_\alpha^{(2)} -
\sum_{s=0}^1\varepsilon^s A_{\alpha s}(x,...,x^{(2s)})\right)=\mathcal{O}(\varepsilon^4).
\end{equation}
Primary constraints due to the possible existence of further null vectors are not
precluded.

(i) We now transform the system \eqref{e:pZB0a} into a more
restrictive set obtained by implying time stability, without
employing the explicit equations of motion yet.

The acceleration dependences in orders higher than $\varepsilon^2$
can be removed from Eq.\ \eqref{e:pZB0a}
by means of the constraints obtained through multiplying the primary constraints
with $\varepsilon$. We are led to
\begin{equation}\label{e:pZB0}
 \varepsilon^2\left(m_\alpha x_\alpha^{(2)} -
\sum_{s=0}^1\varepsilon^s B_{\alpha,2,s}(x, x^{(1)})\right)=\mathcal{O}(\varepsilon^4).
\end{equation}
After differentiating
and eliminating the occurring accelerations by making use of Eqs.
\eqref{e:pZB0}, we find
\begin{equation}\label{e:pZB1}
 \varepsilon^2\left(m_\alpha  x_\alpha^{(3)} -
\sum_{s=0}^1\varepsilon^sB_{\alpha,3,s}(x, x^{(1)})\right)=\mathcal{O}(\varepsilon^4).
\end{equation}
Requiring that the above constraints are satisfied for all times yields the
time stability condition
\begin{equation} \label{e:pZB3}
\varepsilon^2\left(m_\alpha  x_\alpha^{(4)} -
\sum_{s=0}^1\varepsilon^sB_{\alpha,4,s}(x, x^{(1)})\right)=\mathcal{O}(\varepsilon^4).
\end{equation}
 
(ii) Additional constraints emerge from the latter time stability condition 
as well as from the equations
of motion.
They are derived by multiplying
the Euler-Lagrange equations by $\varepsilon$ and thereafter eliminating
all second and higher order time derivatives and found beyond the leading term
with the help of Eqs.\ (\ref{e:pZB0}-\ref{e:pZB3}). A last time
differentiation and elimination of the newly appeared
accelerations leads to the constraints
\begin{equation}
\label{e:pZB4} \varepsilon\left(m_\alpha  x_\alpha^{(r)} -
\sum_{s=0}^2\varepsilon^sB_{\alpha,r,s}(x,
x^{(1)})\right)=\mathcal{O}(\varepsilon^4),\quad r=2,3,
\end{equation}
which are not yet time stable.
The time
stability condition 
\begin{equation}
\label{e:pZB5} \varepsilon\left(m_\alpha  x_\alpha^{(4)} -
\sum_{s=0}^2\varepsilon^sB_{\alpha,4,s}(x, x^{(1)})\right)=\mathcal{O}(\varepsilon^4)
\end{equation}
inserted together with \eqref{e:pZB4} into the equations of motion yields
the more restrictive set of constraints
\begin{equation} \label{e:pZB10}
m_\alpha  x_\alpha^{(r)} - \sum_{s=0}^2\varepsilon^s B_{\alpha,r,s}(x,
x^{(1)})=\mathcal{O}(\varepsilon^4),\quad r=2,3. 
\end{equation}
These constraints are stable, for inserting all constraints plus the time stability
condition into the equations of motion yields a constraint (namely
\eqref{e:pZB10} with $r=2$) that is actually satisfied on the constraint surface.
In contrast to appendix \ref{sec:A},
the time stability condition, and thus the equations of motion had to be employed twice
to determine the constraints of 3pN dynamics.
Finally, we note that
it is now possible to
argue as in Sec.\  \ref{sec:2} that further null vectors of the Hessian do
not lead to further constraints.

\section{\label{sec:C}3pN elementary Dirac brackets}

The issue of this appendix is to present the explicit computation of the Dirac
brackets for the dynamics of interest in this article. The Dirac bracket is
bilinear and acts as a derivation on each argument. To know its action on any two
functions of the phase space, it is therefore sufficient to know all
elementary Dirac brackets, i.e.\ the Dirac brackets of the constraint surface
coordinates. In order to compute them, we have first to determine the Poisson-bracket
matrix $D$ of the constraints up to the needed third order in $\varepsilon$,
appropriately adapting the general method given by \cite{JLM86} to the
3pN case. All expressions reduced to second order in
$\varepsilon$ can be directly used for the post-Coulombian dynamics case.

We start by splitting $D$ into $6\times 6$-submatrices as
\begin{equation}
D=\left(\begin{matrix} S & T \\ -T^T & U
        \end{matrix}\right),
\end{equation}
\begin{align}
S_{\alpha,\beta}&=\{\chi_{1\alpha},\chi_{1\beta}\}
=\varepsilon^2\frac{\pa \Phi_{1\beta}}{\pa x_\alpha^{(1)}}-
\varepsilon^2\frac{\pa \Phi_{1\alpha}}{\pa
x_\beta^{(1)}}+\mathcal{O}\left(\varepsilon^4\right),\\
T_{\alpha,\beta}&=\{\chi_{1\alpha},\omega_{1\beta}\}
= -\delta_{\alpha \beta} -\varepsilon\frac{1}{m_\beta}\frac{\pa
\Phi_{0\beta}}{\pa x_\alpha^{(1)}} +\varepsilon^2\frac{1}{m_\beta}\frac{\pa
\Phi_{1\alpha}}{\pa x_\beta}+\mathcal{O}\left(\varepsilon^4\right),\\
U_{\alpha, \beta}&=\{\omega_{1\alpha},\omega_{1\beta}\}
=\varepsilon\frac{1}{m_\alpha m_\beta}\left(\frac{\pa \Phi_{0 \beta}}{\pa
x_\alpha}-\frac{\pa \Phi_{0 \alpha}}{\pa
x_\beta}\right)+\mathcal{O}\left(\varepsilon^4\right).
\end{align}
This matrix is clearly invertible. To obtain iteratively
the inverse matrix of $D$, we write the expansion in powers of $\varepsilon$
as
\begin{equation}
D=\sum_{s=0}^n\varepsilon^s\,^s\!D+\mathcal{O}(\varepsilon^{n+1}).
\end{equation}
It can be easily verified, that $D^{-1}$ up to
$n$-th order is given by
\begin{equation}
D^{-1}=\left(\mathbf 1+\sum_{s=1}^n \varepsilon^s
\,^s\!N\right)\,^0\!D^{-1}+\mathcal{O}(\varepsilon^{n+1}),
\end{equation}
where $\,^s\!N$ is computed iteratively from
\begin{equation}
\,^s\!N=-\,^0\!D^{-1}\,\,^s\!D-\sum_{r=1}^{s-1}\,^0\!D^{-1}\,\,^{s-r}\!D\,\,^r\!N.
\end{equation}                                               
In our problem, we need $D^{-1}$ up to the third order. Its explicit expression reads
\begin{align}
D^{-1}=&\left(\mathbf
 1+\sum_{s=1}^3\varepsilon^s\,^s\!N\right)\,^0\!D^{-1}+\mathcal{O}(\varepsilon^{4})
\no
 =&\,^0D^{-1}-\varepsilon\, \,^0\!D^{-1}\;\,^1\!D\;\,^0\!D^{-1}
 +\varepsilon^2 (-\,^0\!D^{-1}\;\,^2\!D\;\,^0\!D^{-1}
 +\,^0\!D^{-1}\;\,^1\!D\;\,^0\!D^{-1} \;\,^1\!D\;\,^0\!D^{-1})
\no
 &+\varepsilon^3\left(-\,^0\!D^{-1}\;\,^3\!D\;\,^0\!D^{-1}
  +\,^0\!D^{-1}\;\,^2\!D\;\,^0\!D^{-1} \;\,^1\!D\;\,^0\!D^{-1}
  +\,^0\!D^{-1}\;\,^1\!D\;\,^0\!D^{-1}
 \;\,^2\!D\;\,^0\!D^{-1}
 \right.
\no
 &
\left.\quad -\,^0\!D^{-1}\;\,^1\!D\;\,^0\!D^{-1} \;\,^1\!D\;\,^0\!D^{-1}
 \;\,^1\!D\;\,^0\!D^{-1}\right)+\mathcal{O}(\varepsilon^{4}).
\end{align}
At last, we split $D^{-1}$ into submatrices,
\begin{equation}
D^{-1}=\left(\begin{matrix} X & Y \\ -Y^T & Z
        \end{matrix}\right),
\end{equation}
and account for
\begin{eqnarray*}
\{x_\alpha,\omega_{1\beta}\}&=&\mathcal{O}(\varepsilon^0)\,,\\
\{x_\alpha,\chi_{1\beta}\}&=&\mathcal{O}(\varepsilon^4)\,,\\
\{\Pi_{0\alpha},\omega_{1\beta}\}&=&\mathcal{O}(\varepsilon)\,,\\
\{\Pi_{0\alpha},\chi_{1\beta}\}&=&\mathcal{O}(\varepsilon^2).
\end{eqnarray*}                                               
From the expression \eqref{e:41} for the Dirac bracket 
we conclude directly that, to determine the elementary
Dirac brackets up to the desired third order, we need
$Y$ up to order $1$, $Z$ up to order $2$ and that we do not need $X$ at all.
The components of the relevant submatrices are
\begin{align}
Y_{\alpha,\beta}=&\delta_{\alpha
  \beta}-\varepsilon\frac{1}{m_\alpha}\frac{\pa \Phi_{0\alpha}}{\pa
  x_\beta^{(1)}}+\mathcal{O}(\varepsilon^2)\,,
\\
Z_{\alpha,\beta}=&\varepsilon^2 \left[\frac{\pa
\Phi_{1\beta}}{\pa x_\alpha^{(1)}}-\frac{\pa \Phi_{1\alpha}}{\pa
x_\beta^{(1)}}\right]
  -\varepsilon^3 \left[\sum_\gamma\left( \frac{\pa
\Phi_{1\gamma}}{\pa x_\alpha^{(1)}}-\frac{\pa \Phi_{1\alpha}}{\pa
x_\gamma^{(1)}}\right) \left(\frac{1}{m_\gamma}\frac{\pa \Phi_{0\gamma}}{\pa
x_\beta^{(1)}}\right) \right.
\no
&
\left.
-\left(\frac{\pa \Phi_{1\gamma}}{\pa
x_\beta^{(1)}}-\frac{\pa \Phi_{1\beta}}{\pa x_\gamma^{(1)}}\right)
\left(\frac{1}{m_\gamma}\frac{\pa \Phi_{0\gamma}}{\pa
x_\alpha^{(1)}}\right)\right]+\mathcal{O}(\varepsilon^4).
\end{align}
Hence the elementary Dirac brackets read
\begin{align}\label{e:EDB1}
\{x_\alpha,x_\beta\}^*=&\frac{1}{m_\alpha
m_\beta}Z_{\alpha,\beta}+\mathcal{O}(\varepsilon^4)\,,
\\
\{x_\alpha,\Pi_\beta\}^*=&\delta_{\alpha \beta}
+\varepsilon^2\frac{1}{m_\alpha}\frac{\pa \Phi_{1 \alpha}}{\pa x_\beta}
\no&
+\varepsilon^3\sum_\gamma \frac{1}{m_\alpha m_\gamma} 
\bigg[ \bigg(\frac{\pa
\Phi_{1 \gamma}}{\pa x_\alpha^{(1)}}-\frac{\pa \Phi_{1 \alpha}}{\pa
x_\gamma^{(1)}}\bigg) \frac{\pa \Phi_{0\gamma}}{\pa x_\beta}
-
 \frac{\pa\Phi_{0\gamma}}{\pa x_\alpha^{(1)}} \frac{\pa \Phi_{1\gamma}}{\pa
x_\beta}\bigg]+\mathcal{O}(\varepsilon^4)\,,
\\ 
\label{e:EDB3}
\{\Pi_\alpha,\Pi_\beta\}^*=&\epsilon^3
\sum_\gamma\frac{1}{m_\gamma}\left(\frac{\pa \Phi_{0\gamma}}{\pa x_\alpha}
\frac{\pa \Phi_{1\gamma}}{\pa x_\beta}- \frac{\pa \Phi_{0\gamma}}{\pa x_\beta}
\frac{\pa \Phi_{1\gamma}}{\pa x_\alpha}\right)+\mathcal{O}(\varepsilon^4).
 \end{align}   
Finally, we explicitly display the relation \eqref{e:EDB1} up to second order
for both 3pN and 2pC dynamics.
 \begin{align}
\text{$2$pN:}\quad \{x_{ai},
x_{bj}\}^*=&\varepsilon^2\frac{G}{4}\Bigg[7\left(\frac{\Pi_{0\,ai}}{m_a}+
\frac{\Pi_{0\,bi}}{m_b}\right)n_{ab\,j}
-7\left(\frac{\Pi_{0\,aj}}{m_a}+\frac{\Pi_{0\,bj}}{m_b}\right)n_{ab\,i}
\no
&+\left(7\delta_{ij}+n_{ab\,i}n_{ab\,j}\right)\left(\sum_{k=1}^3
n_{ab\,k}\left(\frac{\Pi_{0\,ak}}{m_a}+\frac{\Pi_{0\,bk}}{m_b}\right)\right)
\Bigg]\,,
\label{e:C19}
\\
\text{$2$pC:}\quad \{x_{ai},
x_{bj}\}^*=&-\varepsilon^2\frac{e_ae_b}{m_am_b}
\Bigg[\left(\frac{\Pi_{0\,ai}}{m_a}+\frac{\Pi_{0\,bi}}{m_b}\right)n_{ab\,j}
-\left(\frac{\Pi_{0\,aj}}{m_a}+\frac{\Pi_{0\,bj}}{m_b}\right)n_{ab\,i}
\no
&
+\left(\delta_{ij}+n_{ab\,i}n_{ab\,j}\right)\left(\sum_{k=1}^3
n_{ab\,k}\left(\frac{\Pi_{0\,ak}}{m_a}+\frac{\Pi_{0\,bk}}{m_b}\right)\right)\Bigg].
\label{e:C20}
 \end{align}

\bibliographystyle{apsrev}
\bibliography{literature}

\end{document}